\documentclass[a4paper,amsmath,amssymb,superscriptaddress,aps,prb,reprint,showpacs]{revtex4-2}
\usepackage{xcolor,graphicx}
\usepackage{subfigure}
\usepackage{MnSymbol}
\usepackage{braket}
\usepackage{tikz}
\usepackage{pgfplots}
\usepackage[utf8]{inputenc}
\usepackage{floatrow}
\newfloatcommand{capbtabbox}{table}[][\FBwidth]

\renewcommand{\onlinecite}[1]{\cite{#1}}
\DeclareMathOperator*{\Motimes}{\text{\raisebox{0.25ex}{\scalebox{0.8}{$\bigotimes$}}}}
\begin{document}

\title{Magnetic phases for two $t_{2g}$ holes with spin-orbit coupling and crystal field}

\author{Pascal Strobel}
\author{Friedemann Aust}
\affiliation{%
Institut f\"ur Funktionelle Materie und Quantentechnologien,
Universit\"at Stuttgart,
70550 Stuttgart,
Germany}

\author{Maria Daghofer}
\affiliation{%
Institut f\"ur Funktionelle Materie und Quantentechnologien,
Universit\"at Stuttgart,
70550 Stuttgart,
Germany}
\affiliation{Center for Integrated Quantum Science and Technology, University of Stuttgart,
Pfaffenwaldring 57, 
70550 Stuttgart, Germany}

\date{\today}

\begin{abstract}
We investigate two holes in the the $t_{2g}$
levels of a square-lattice Mott insulator with strong spin-orbit
coupling. Exact diagonalization of a spin-orbital model valid at
strong onsite interactions, but arbitrary spin-orbit coupling and
crystal field is complemented by an effective triplon model (valid for
strong spin-orbit coupling) and by a semiclassical variant of the
model. We provide the magnetic phase diagram depending on crystal
field and spin-orbit coupling, which largely agrees for
the semiclassical and quantum models, as well as excitation spectra
characterizing 
the various phases. 
\end{abstract}

\maketitle

\section{Introduction}  \label{sec:intro}
The interplay between spin-orbit coupling (SOC) and correlated electrons as a driving force of physical properties in transition metal compounds has gathered significant interest in the last decade \cite{doi:10.1146/annurev-conmatphys-020911-125138,doi:10.1146/annurev-conmatphys-031115-011319}. The manifold of competing interactions in these materials has led to a plethora of interesting properties like topological Mott insulators, superconductivity and spin liquids \cite{Pesin:2010ju}.

Focus was first on materials with one hole in the $t_{2g}$ manifold
and strong SOC in addition to sizable correlations, as realized in $4d$ and $5d$ states. SOC couples spin $\textbf{S}=1/2$ and orbital $\textbf{L}=1$ degrees of freedom to a total angular moment $\textbf{J}=1/2$, so that the model in the end can be described by an effective half-filled model. In addition to similarities to high-$T_C$ cuprates and the potential realization \cite{Chaloupka:2010gi} of the exactly solvable Kitaev model \cite{Kitaev:2006ik} in a honeycomb lattice, which have stimulated extensive research on these compounds \cite{rev_square_iri,0953-8984-29-49-493002}, potential applications in spintronics have been proposed more recently~\cite{Sr2IrO4memory}. 

Interest was then extended to other
fillings~\cite{PhysRevLett.118.086401,PhysRevB.98.205128}, and we will
here focus on the Mott-insulating state for two holes. For dominant
SOC (as possibly in Ir), the system is in the $j$-$j$ limit and the
groundstate is thus likely a nonmagnetic ground
state~\cite{PhysRevB.93.035129,PhysRevLett.120.237204} given by two
holes filling the $j=1/2$ states. For weaker SOC, e.g. in
ruthenates,  $\textbf{L}$-$\textbf{S}$ coupling is more appropriate,
where SOC couples $\textbf{L}=1$ and $\textbf{S}=1$ to $\textbf{J}=0$,
again  leading to a nonmagnetic ground state for a single
ion~\cite{RIXS_Ca2RuO4_Gretarson19}. However, energy scales are here
rather different with a much smaller splitting between the singlet and
triplet states. When going from an isolated ion to a compound with a
lattice, competing processes can overcome the splitting. Superexchange
mixes in states from the $\textbf{J}=1$ level, which can lead to a
magnetic ground state. 

This phenomenon is also known as excitonic or Van-Vleck
magnetism~\cite{PhysRevLett.111.197201}, and has for instance been
proposed to provide a route to a bosonic Kitaev-Heisenberg model
\cite{PhysRevLett.122.177201,PhysRevB.100.224413} and to explain
magnetic excitations of $\text{Ca}_2\text{RuO}_4$~\cite{Higgs_Ru}. In
one dimension, density-matrix renormalization group has been applied
to a spin-orbit coupled and correlated $t_{2g}$ model with two holes,
and antiferromagnetic (AFM) order has been
found~\cite{PhysRevB.96.155111,PhysRevB.101.245147} both for
intermediate correlations (of a more 'standard' excitonic type with
intersite pairs) and for strong correlations (of the 'onsite' type
discussed in Ref.~\onlinecite{PhysRevLett.111.197201}). Similarly,
dynamical mean-field theory has yielded excitonic antiferromagnetism
in a two-dimensional model~\cite{PhysRevB.99.075117}. 

A material which has been a focal point of discussions in this context
is $\text{Ca}_2\text{RuO}_4$. In
neutron scattering experiments an in-plane AFM ordering has been
measured below the N\'eel temperature $T_N\approx 110\,K$ and
neutron-scattering spectra can only be explained by taking into
account substantial
SOC~\cite{PhysRevLett.119.067201,PhysRevLett.95.136401,PhysRevLett.87.077202,PhysRevLett.115.247201}. Accordingly,
excitonic magnetism, where the 
magnetic moment arises from admixture of $\textbf{J}=1$ component into
the ionic $\textbf{J}=0$ state, has been argued to describe this
compound~\cite{PhysRevLett.119.067201,Higgs_Ru}.  However, a strong
crystal field (CF), favoring doubly occupied $xy$ orbitals, is also
clearly present in Ca$_2$RuO$_4$ and complicates the analysis, because
it would favor a description in terms of a spin-one system. This is
backed by a structural phase transition accompanying the
metal-insulator transition. SOC would in this picture be only  a
correction affecting
excitations~\cite{PhysRevLett.115.247201,PhysRevB.101.205128}.  

In a previous publication, some of us have used the variational
cluster approach (VCA) based on {\it ab initio} parameters to show that
excitonic antiferromagnetism can coexist with substantial CF's and that
Ca$_2$RuO$_4$ falls into this
regime~\cite{PhysRevResearch.2.033201,2021arXiv210205489L} of orbitally polarized
excitonic antiferromagnetism. In the present paper, we
study the competition of CF $\Delta$ and SOC $\lambda$ in
$t_{2g}^4$ systems in more depth and for a wider parameter space. We
investigate an effective spin-orbit model obtained in second-order
perturbation theory, as also used for
$\text{Ca}_2\text{RuO}_4$~\cite{PhysRevResearch.2.033201}. This 
extends the comparison of CF and SOC acting on the itinerant  
regime (without magnetic ordering)~\cite{PhysRevB.98.205128} to
magnetic Mott insulators. Our work is also complementary to a very recent study using the Hartree-Fock approach to investigate the dependence of magnetic ordering on SOC, CF, and tilting of octahedra, which focused on patterns with smaller unit cells of one or two Ca ions~\cite{Mohapatra_2020}. We obtain
$\Delta$-$\lambda$ phase diagrams using both Monte-Carlo (MC) simulations
for the semiclassical limit of the model and exact diagonalization (ED) for the
quantum system and provide excitation spectra for the various magnetic
phases.

As expected~\cite{PhysRevResearch.2.033201}, stripy magnetism is found when both SOC and CF are weak,
and checkerboard order (as seen in Ca$_2$RuO$_4$) takes over when
either becomes strong enough to sufficiently lift orbital
degeneracy. For negative CF, i.e., disfavoring doubly occupied $xy$
orbitals, we find an additional intermediate phase with rather complex
magnetic order. Overall, we find the agreement between the semiclassical and
quantum models to be quite good, with phase boundaries between the
magnetic phases only moderately different.
Similarly, the transition
to a paramagnetic (PM) state at strong SOC in the full
quantum-mechanical model is compared to an  effective triplon
model~\cite{PhysRevLett.111.197201}, valid at strong SOC, and found to
agree.
Finally, we present the dynamic spin structure factor of the
spin-orbital model to discuss signatures of the various magnetic
phases accessible to neutron scattering experiments.

In Sec.~\ref{sec:model}, we introduce models, i.e., the full
spin-orbital superexchange model as well as the triplon model valid for strong SOC,
and methods. In
Sec.~\ref{sec:LR}, we first go over the limiting cases of the spin-orbital system
at dominant CF, the triplon scenario, discuss
the intricate interplay of spin and orbital order for small CF and
SOC, and finally give the phase diagram for intermediate values in
Sec.~\ref{sec:phase_diag}. The phase diagram is compared to results of
semiclassical MC calculations for the same model in
Sec.~\ref{sec:MCMC}.  Section~\ref{sec:Skw} presents the dynamic-spin-structure-factor data corresponding to neutron scattering
experiments for the various phases. 
Finally Sec.~\ref{sec:conclusions} gives a summary of the results found in this paper.

\section{Model and Methods}\label{sec:model} 

\subsection{Spin-orbit model}\label{sec:spinorbit}
\begin{figure}
\begin{floatrow}
\ffigbox[0.5\textwidth]{%
 \includegraphics[width=1.1\columnwidth]{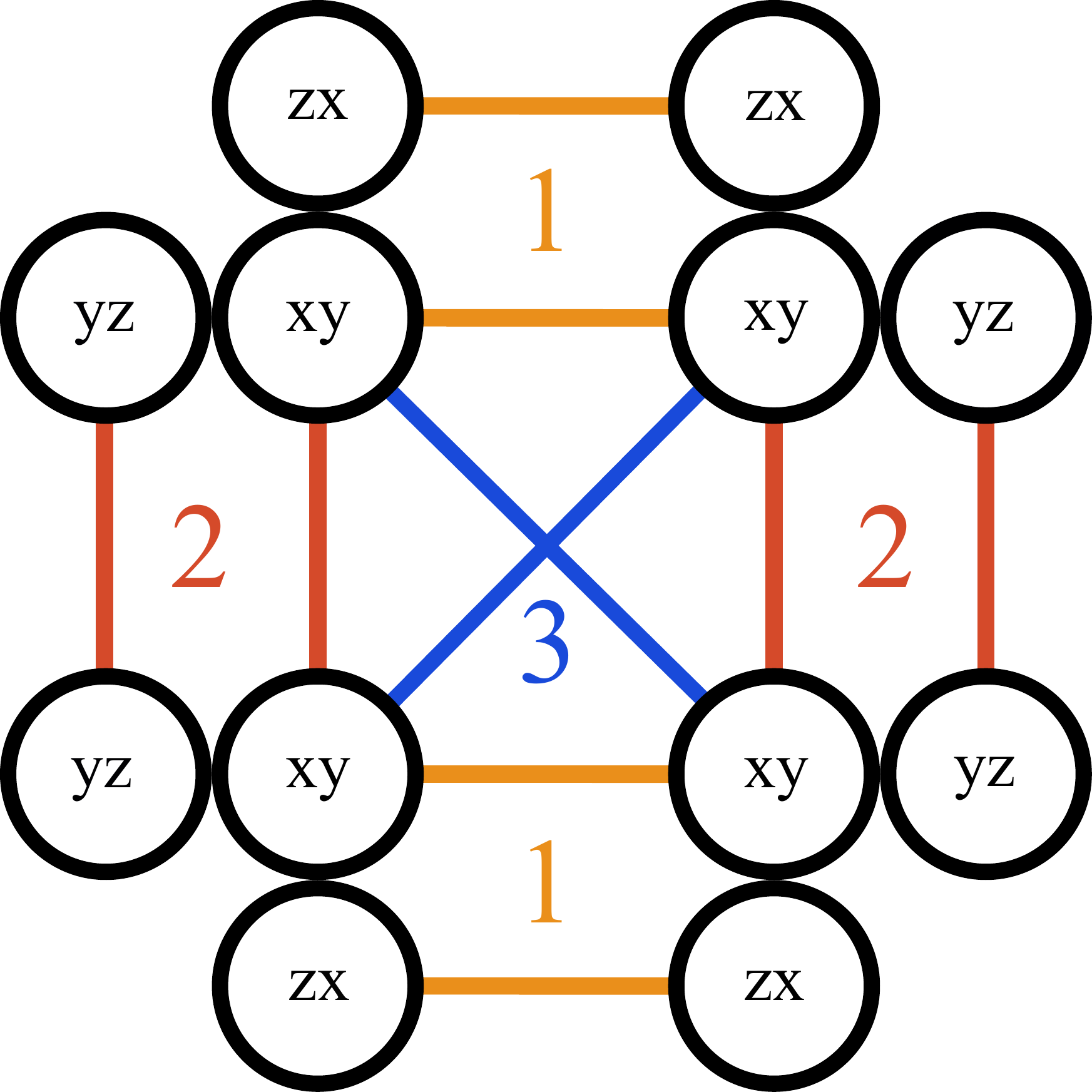}%
}{ \caption{Possible hopping processes in $\text{Ca}_2\text{RuO}_4$ (based on \cite{PhysRevResearch.2.033201}). The $xy$ orbital can hop in $x$- (bond 1) and $y$-direction (bond 2) and has a nonzero hopping amplitude for next-nearest neighbors (bond 3). The $zx$ and $yz$ orbital can hop only on bond 1 and 2 respectively.   \label{fig:Fig1}}}
\capbtabbox[0.43\textwidth]{%
  \begin{tabular}{cc} \hline
  $t_{\alpha,m}$ & Amplitude \\ \hline
 $t_{xy,1}$ & $t_{xy}$ \\
 $t_{xy,2}$& $t_{xy}$\\
 $t_{xy,3}$& $t_{\text{NNN}}$\\
 $t_{zx,1}$& $t_{zx}$\\
 $t_{zx,2}$& $0$\\
 $t_{zx,3}$& $0$\\
 $t_{yz,1}$& $0$\\
 $t_{yz,2}$& $t_{yz}$\\
 $t_{yz,3}$& $0$\\ \hline
  \end{tabular}
}{\caption{Possible hopping parameters $t_{\alpha,m}$ from equations (\ref{eq:Eq1}) and (\ref{eq:Eq4})-(\ref{eq:Eq6}) as well as their amplitudes for a square lattice geometry. The parameter $m$ here indicates the bond type introduced in Fig.\ref{fig:Fig1} while $\alpha$ are the $t_{2g}$ orbitals.\label{tab:TabI}}}
\end{floatrow}
 
\end{figure}

\begin{figure}
\includegraphics[width=\columnwidth]{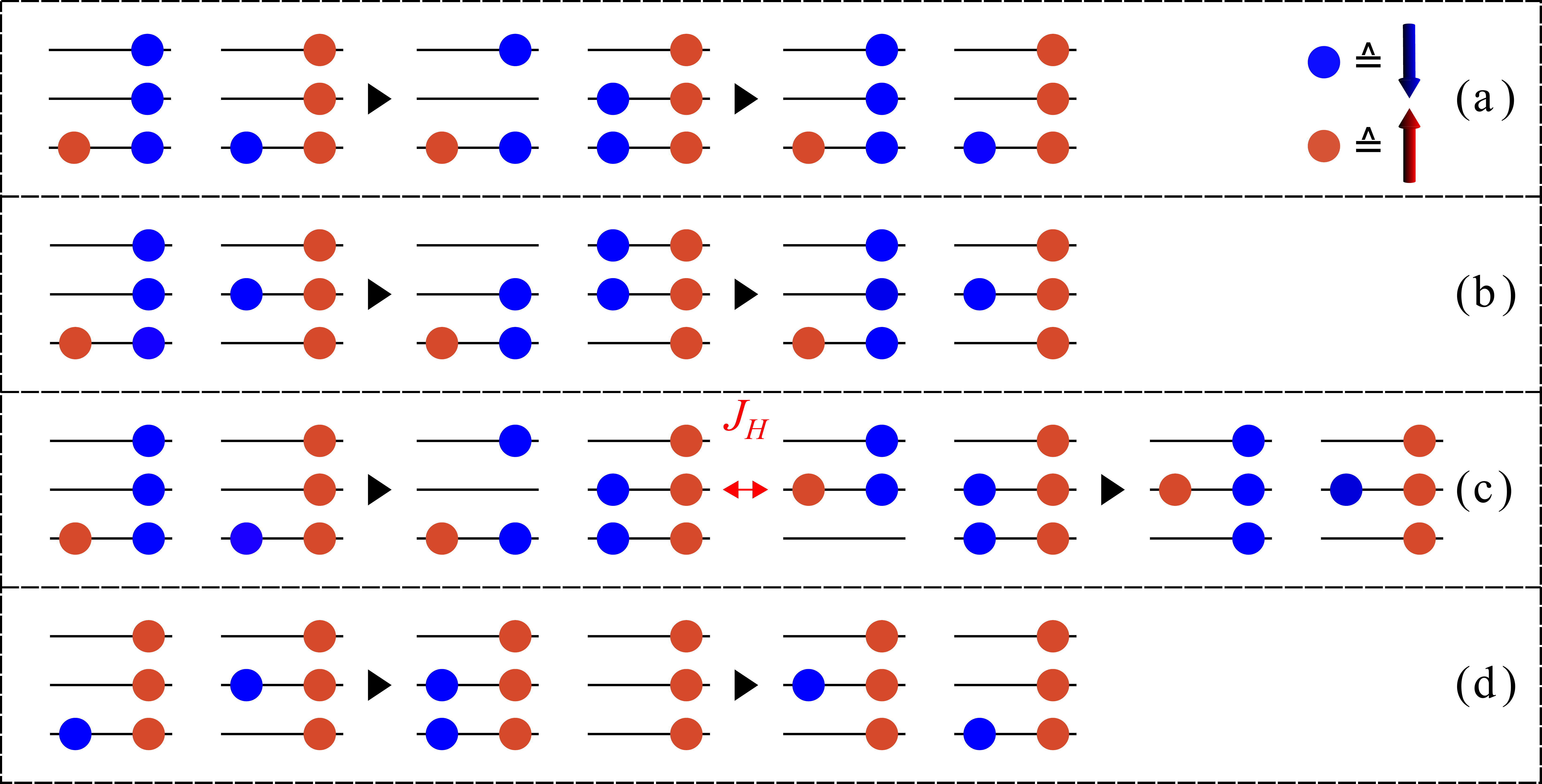}
\caption{Displayed are the different possible hopping processes from (\ref{eq:Eq5}) and (\ref{eq:Eq4}). In (a) and (b) virtual hoppings where the orbital configuration is preserved are shown. In (a) the double occupancy is at the same orbital, while in (b) the double occupancy resides at different orbitals. (c) and (d) display second order hoppings where the orbital configurations change. In the ``pair-flip'' process (c) the change arises due to the last term in (\ref{eq:Eq2}), while the ``swap'' process (d) arises due to a different orbital hopping back then forth.}  \label{fig:Fig2}
\end{figure}

$\text{Ca}_2\text{RuO}_4$ has a $d^4$ configuration, meaning that four electrons reside in three $t_{2g}$ orbitals, from now on referred to as $xy$-, $zx$-, and $yz$-orbital. The kinetic part of this Hamiltonian can be written as 
\begin{align}\label{eq:Eq1}
H_{\text{kin}}&=\sum_{m=1}^{3}\sum_{\braket{i,j}_m}\sum_{\alpha,\sigma}(t_{\alpha,m}c_{i,\alpha,\sigma}^{\dagger}c_{j,\alpha,\sigma} + h.c.),
\end{align}
where $m$ are the three different bond types introduced in Fig.~\ref{fig:Fig1} and $t_{\alpha,m}$ is the hopping amplitude depending on the orbital flavor $\alpha$ and the bond type $m$. Tab.~\ref{tab:TabI} gives the amplitudes for all possible $t_{\alpha,m}$ for a square lattice geometry. 
$c_{i,\alpha,\sigma}^{\dagger}$ ($c_{i,\alpha,\sigma}$) is creating (annihilating) an electron in orbital $\alpha$ at site $i$ with spin $\sigma$. The possible hopping paths for $\text{Ca}_2\text{RuO}_4$ \cite{PhysRevResearch.2.033201,PhysRevB.95.014409} are shown in Fig.~\ref{fig:Fig1}. On nearest neighbor bonds (NN) only two orbitals are active (e.g. $xy$ and $zx$ for $x$-bonds), while for next-nearest neighbor bonds (NNN) only the $xy$ orbital has a nonzero hopping amplitude (see Tab.~\ref{tab:TabI}). 

The onsite interaction has the form of a Kanamori-Hamiltonian \cite{PhysRevB.28.327}
\begin{align}\label{eq:Eq2}
H_{\text{int}}=&U\sum_{i,\alpha}n_{i\alpha\uparrow}n_{i\alpha\downarrow}+U'\sum_{i,\sigma}\sum_{\alpha<\beta}n_{i\alpha\sigma}n_{i\beta \,-\sigma}\notag\\
   &+(U'-J_H)\sum_{i,\sigma}\sum_{\alpha<\beta}n_{i\alpha\sigma}n_{i\beta\sigma}\notag\\
   &-J_H\sum_{i,\alpha\neq\beta}(c_{i\alpha\uparrow}^{\dagger}c_{i\alpha\downarrow}c_{i\beta\downarrow}^{\dagger}c_{i\beta\uparrow}-c_{i\alpha\uparrow}^{\dagger}c_{i\alpha\downarrow}^{\dagger}c_{i\beta\downarrow}c_{i\beta\uparrow}),
\end{align}
with intraorbital Hubbard interaction $U$, interorbital $U'=U-2J_H$ and Hund's coupling $J_H$. 

Since the computational cost to calculate this Hamiltonian via ED is very high we only consider a low energy sector
of our Hilbert space. We focus here on the Mott-insulating regime with
large $U$ and $J_H$. The low-energy sector is then given by states
where each site contains exactly four electrons (two holes), as $U$
suppresses charge fluctuations. Hund's-rule coupling $J_H$ moreover
ensures that exactly one orbital per site is
doubly occupied and that the electrons in the remaining two
half-filled orbitals form a total spin $\mathbf{S}=1$. This means we
have three different orbital configurations and a $\mathbf{S}=1$ spin state,
leading to a subspace of nine states. The orbital
configurations are labeled with the orbital which is doubly occupied
from here on. It turns out (see \cite{PhysRevLett.105.027204}) that
this orbital degree of freedom can be mapped to an effective angular
momentum with   
\begin{align}\label{eq:Eq3}
L^{x}=\mathcal{L}_{yz}=-\text{i}(\ket{xy}\bra{zx}-\ket{zx}\bra{xy})\notag\\
L^{y}=\mathcal{L}_{xz}=-\text{i}(\ket{yz}\bra{xy}-\ket{xy}\bra{yz})\notag\\
L^{z}=\mathcal{L}_{xy}=-\text{i}(\ket{zx}\bra{yz}-\ket{yz}\bra{zx}),
\end{align}
where the notation $\mathcal{L}_{\alpha}$ with an orbital index $\alpha$ is introduced to make the expression of equations (\ref{eq:Eq4})-(\ref{eq:Eq6}) more straightforward and can be easily translated into the $x$-, $y$- and $z$-component of the angular momentum $\mathbf{L}$.

The effective spin-orbital Hamiltonian is then obtained by treating
the hopping term in second order perturbation theory. This gives 
a Kugel-Khomskii type Hamiltonian \cite{Streltsov,Kugel}, where only
virtual hopping processes of the form $d^4d^4\rightarrow
d^5d^3\rightarrow d^4d^4$ take place.
The effective spin-orbital superexchange Hamiltonian includes both orbital as well as spin-orbital interactions. 
Spin-orbital superexchange terms that preserve orbital occupations of the two sites are
\begin{align}\label{eq:Eq4}
H_{\text{OP}}=&\sum_{m=1}^{3}\sum_{\braket{i,j}_m}\sum_{\alpha\neq\beta}\bigg[t_{\beta,m}^2\frac{U+J_H}{U(U+2J_H)}\notag\\
&\times(\mathbf{S}_i\mathbf{S}_j-1)(1-\mathcal{L}_{\alpha}^2)_i(1-\mathcal{L}_{\alpha}^2)_j\notag\\
&+\bigg(t_{\gamma\neq(\alpha,\beta),m}^2\frac{(U+J_H)}{U(U+2J_H)}
 -\frac{(t_{\alpha,m}^2+t_{\beta,m}^2)J_H}{U(U-3J_H)}\bigg)\notag\\
 &\times(\mathbf{S}_i\mathbf{S}_j-1)(1-\mathcal{L}_{\alpha}^2)_i(1-\mathcal{L}_{\beta}^2)_j\bigg].
\end{align}
Here we used the aforementioned mapping from orbitals to effective angular momentum $\mathbf{L}$.
Having two orbitals of the same flavor means only the electrons in the other two orbitals are allowed to perform a virtual hopping (Fig.\ref{fig:Fig2} (a)), while for different flavors each orbital can be involved in such a hopping process (Fig.\ref{fig:Fig2} (b)).

Furthermore, there are spin-orbital couplings that change orbital
configurations. These can be separated in so called ``pair-flip'' (Fig.\ref{fig:Fig2} (c)) processes where two orbitals of the same flavor flip their flavor to another one and ``swap'' processes (Fig.\ref{fig:Fig2} (d)) where two orbitals of different flavor exchange their flavor
\begin{align}\label{eq:Eq5}
H_{\text{OF}}=&\sum_{m=1}^{3}\sum_{\braket{i,j}_{m}}\sum_{\alpha\neq\beta}\bigg[-t_{\alpha,m}t_{\beta,m}\frac{J_H}{U(U+2J_H)}\notag\\
&\times(\mathbf{S}_i\mathbf{S}_j-1)(\mathcal{L}_{\beta}\mathcal{L}_{\alpha})_i(\mathcal{L}_{\beta}\mathcal{L}_{\alpha})_j\notag\\
&+\bigg(t_{\alpha,m}t_{\beta,m}\frac{(U-J_H)}{U(U-3J_H)}\bigg)\notag\\
&\times(\mathbf{S}_i\mathbf{S}_j+1)(\mathcal{L}_{\beta}\mathcal{L}_{\alpha})_i(\mathcal{L}_{\alpha}\mathcal{L}_{\beta})_j\bigg].
\end{align}
Finally, additional orbital terms affect sites $i$ and $j$ with
different  orbital occupation:
\begin{align}\label{eq:Eq6}
H_{\mathbf{L}\cdot\mathbf{L}}=&\sum_{m=1}^{3}\sum_{\braket{i,j}_m}\sum_{\alpha\neq\beta}\bigg[t_{\alpha,m}t_{\beta,m}\frac{2J_H}{U(U-3J_H)}\notag\\
&\times(\mathcal{L}_{\beta}\mathcal{L}_{\alpha})_i(\mathcal{L}_{\alpha}\mathcal{L}_{\beta})_j\notag\\
&-(t_{\alpha,m}^2+t_{\beta,m}^2)\frac{1}{(U-3J_H)}\notag\\
&\times(1-\mathcal{L}_{\alpha}^2)_i(1-\mathcal{L}_{\beta}^2)_j\bigg].
\end{align}
The full superexchange interaction of two sites can be summarized as
\begin{align}
H=H_{\text{OF}}+H_{\text{OP}}+H_{\mathbf{L}\cdot\mathbf{L}}
\end{align}
Using the hoppings symmetry allowed on a square lattice up to
second neighbors (see Fig.~\ref{fig:Fig1} and Tab.~\ref{tab:TabI}), one obtains the
effective spin-orbital model that can, e.g., be applied to
$\text{Ca}_2\text{RuO}_4$~\cite{PhysRevResearch.2.033201}. 

In addition to these intersite interactions we also include SOC
$\lambda$ and the CF splitting $\Delta$. The SOC terms can be written in the form
\begin{align}
H_{\text{SOC}}=\lambda\sum_i\mathbf{S}_i\cdot\mathbf{L}_i=\text{i}\lambda\sum_i\sum_{\substack{\alpha,\beta,\gamma\\ \sigma,\sigma'}}\epsilon_{\alpha\beta\gamma}\tau^{\alpha}_{\sigma\sigma'}c^{\dagger}_{i,\beta,\sigma}c_{i,\gamma,\sigma'}, 
\end{align}
where $\epsilon_{\alpha\beta\gamma}$ denotes the Levi-Civita symbol and $\tau^{\alpha}$ are Pauli matrices~\cite{PhysRevB.98.205128,PhysRevB.73.094428}. SOC favors the total angular momentum to be $J=0$, while the CF favors a double occupancy of the $xy$
orbital. Projected onto the low-energy Hilbert space spanned by
$\mathbf{S}=1$ and $\mathbf{L}=1$, they can be written as
\begin{align}
H_{\text{Ion}}=H_{\text{SOC}}+H_{\text{CF}}=\lambda\sum_i\mathbf{S}_i\mathbf{L}_i+\Delta \sum_i(L^z_i)^2.
\end{align}
Going beyond previous effective
models~\cite{PhysRevLett.111.197201,PhysRevB.90.035137,PhysRevLett.102.017205}, 
our model thus fully captures the influence of the Hund's coupling $J_H$ and
gives the possibility to investigate anisotropic hoppings as well as the
$\lambda,\Delta\rightarrow 0$ limits.

The competition between the last two terms, CF $\Delta$ and SOC
$\lambda$, is one of the main topics of this paper. We thus fix the
remaining parameters to values appropriate for
Ca$_2$RuO$_4$~\cite{PhysRevLett.123.137204}. Hopping processes between NN sites and NNN sites were
included with hopping parameters set
to $t_{xy}=0.2\,\text{eV}$, $t_{yz}=t_{zx}=0.137\,\text{eV}$, $t_{\text{NNN}}=0.1\,\text{eV}$,
and $\Delta=0.25\,\text{eV}$ via density-functional theory~\cite{PhysRevLett.123.137204}. However, we found that results only
differ in details when more symmetric NN hoppings
$t_{xy}=t_{yz}=t_{zx}$ are used or when NNN hopping is left off. 
Substantial onsite Coulomb repulsion and Hund's-rule
coupling $U=2\,\text{eV}$ and $J_H=0.34\,\text{eV}$, as can be
inferred from x-ray studies \cite{PhysRevB.100.045123}, 
stabilize a Mott
insulator with robust onsite spin $\mathbf{S}=1$. Previous
calculations using the VCA have shown~\cite{PhysRevResearch.2.033201} that most of the weight of the ground state is indeed captured
by states that minimize Coulomb interactions (\ref{eq:Eq2}), so that a
super-exchange treatment and the resulting spin-orbital model can be justified.

\subsection{PM phase and triplon model}\label{sec:triplons}

\begin{figure}
\includegraphics[width=\columnwidth]{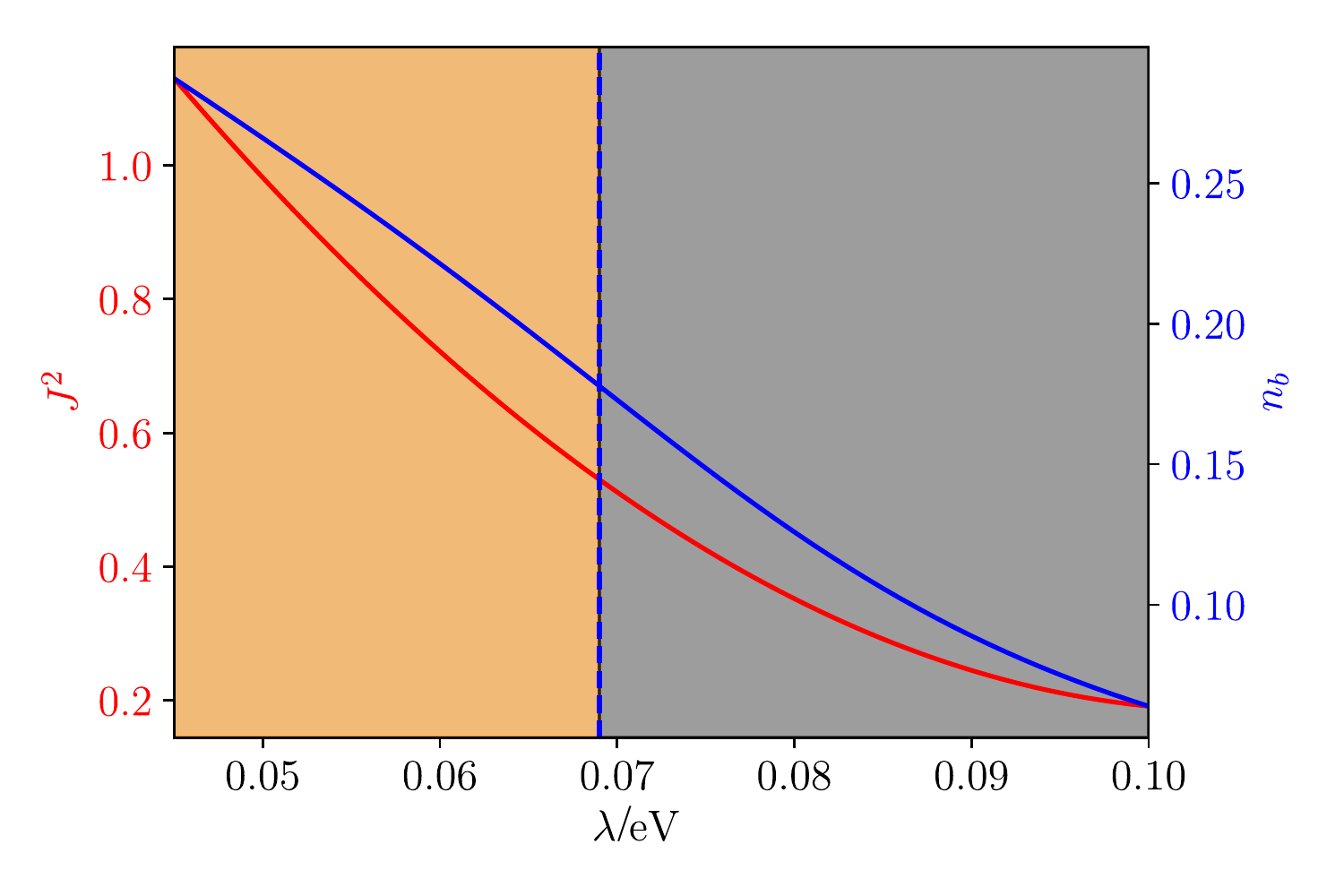}
\caption{$\braket{J^2}$ of the spin-orbit model (red) and triplon
  number $n_b$ of the triplon model (blue) are plotted in dependency of
  SOC $\lambda$. The dashed blue line denotes the phase transition to the PM phase in the triplon model, which is determined via $\frac{d^2n_b}{d\lambda^2}=0$. The parameters are chosen to be $t_{xy}=0.2\,\text{eV},\,t_{yz}=t_{zx}=0.137\,\text{eV},\,t_{\text{NNN}}=0.1\,\text{eV}$, $U=2\,\text{eV}$, $J_H=0.34\,\text{eV}$, and $\Delta=0.1\,\text{eV}$. \label{fig:Fig3}
}
\end{figure}

For strong SOC, we expect our system to be in a PM phase
where each ion is in the $J=0$
state~\cite{PhysRevLett.111.197201,PhysRevResearch.2.033201}. Transition
into magnetically ordered states occurs then via condensation of
triplons. We are going to compare the large-SOC limit of the full
spin-orbit superexchange model to a triplon model appropriate for
significant SOC. We take an approach like in
Ref.~\cite{PhysRevLett.102.017205} and project (\ref{eq:Eq4})-(\ref{eq:Eq6})
onto the low energy subspace of the SOC Hamiltonian, i.e., onto the  $J=0$ and
$J=1$ states
\begin{align}
&\ket{J=0,M_J=0}=\frac{1}{\sqrt{3}}(\ket{M_S=1,M_L=-1}+\ket{-1,1}-\ket{0,0})\notag\\
&\ket{J=1,M_J=1}=\frac{1}{\sqrt{2}}(\ket{1,0}-\ket{0,1})\notag\\
&\ket{J=1,M_J=0}=\frac{1}{\sqrt{2}}(\ket{1,-1}-\ket{-1,1})\notag\\
&\ket{J=1,M_J=-1}=\frac{1}{\sqrt{2}}(\ket{-1,0}-\ket{0,-1})
\end{align}
and projecting out the  $J=2$ levels.

We then can define triplon operators  $T_{1/0/-1}^{\dagger}$
($T_{1/0/-1}$) which  create (annihilate) the respective $J=1$ triplet
state and annihilate (create) the $J=0$ singlet. These operators can
then be rewritten to $T_{x/y/z}$ (for further details see
\cite{PhysRevLett.111.197201}).

\subsection{Methods}\label{sec:methods} 
To investigate these models we use ED on an eight site cluster with 
$\sqrt{8}\times\sqrt{8}$ geometry to determine a $\Delta-\lambda$ phase diagram as well as the dynamical spin structure factor (DSSF) for specific $\Delta$ and $\lambda$.

This is done for both the full spin-orbital model (Sec.~\ref{sec:spinorbit}) as well as the triplon model introduced in Sec.~\ref{sec:triplons}. While the spin-orbital model is capable of capturing the physics at weak SOC, for strong SOC the triplon model is numerically more accessible due to the reduction of the Hilbert space.
 
To confirm the results of ED and get a better understanding of the
phases identified, we additionally performed semiclassical parallel
tempering MC calculations with the full spin-orbital model
for a $4\times 4$ cluster. The easier approach of a fully classical treatment, meaning a parametrization of $\textbf{S}_i$ and $\textbf{L}_i$ as three dimensional real vectors, is not sufficient here. A simple example can be found in the $(\mathcal{L}_{\beta}\mathcal{L}_{\alpha})_i$ terms in the Hamiltonian. There is a clear difference between calculating this expression with scalar components of a three dimensional vector and representing the angular momenta as non-commutative matrices. We accomplish the latter by instead considering trial wave functions of direct-product form 
\begin{align}
\ket{\Psi}=\Motimes_i \left(\ket{S_i} \otimes \ket{L_i} \right)\;, 
\end{align}
where the first product runs over all sites $i$. We allow all complex linear combinations of the $L^z$ eigenvalues $\ket{L_i}=\mu_{1,i} \ket{M_L=-1}+ \mu_{2,i} \ket{M_L=0} + \mu_{3,i} \ket{M_L=+1}$ with $\boldsymbol{\mu}_i^T \boldsymbol{\mu}_i^*=1$, and analogously for the spin $\ket{S_i}$. These trial wave functions are used to calculate the energy, i.e., the energy becomes a (real-valued) function of classical complex vectors $\boldsymbol{\mu}_i$. Classical Markov-chain Monte Carlo is then based on this energy function.

 A similar approach has been used for quadrupole correlations in a spin-1 model with biquadratic interaction ~\cite{PhysRevB.79.214436}. In this context one might refer to our method as a $SU(3) \otimes SU(3)$ semiclassical Monte-Carlo simulation. Compared to ED the numerical expenses of this method are minute. A big drawback of the product state nature of the basis is its inability to accurately represent the singlet and hence find the paramagnetic phase. However, we have the triplon model to confirm ED data in this parameter range. The Monte Carlo code is used as a counterpart of the triplon model for low spin-orbit coupling. 
 
  Finally we point out that all terms in the Hamiltonian are represented as matrices in the chosen basis and the scalar definitions of spin components or other observables are recovered by simply constructing the expectation values regarding $\ket{\Psi}$.

\section{Results}\label{sec:results} 

\subsection{Limiting regimes}\label{sec:LR}

\begin{figure}
\includegraphics[width=\columnwidth]{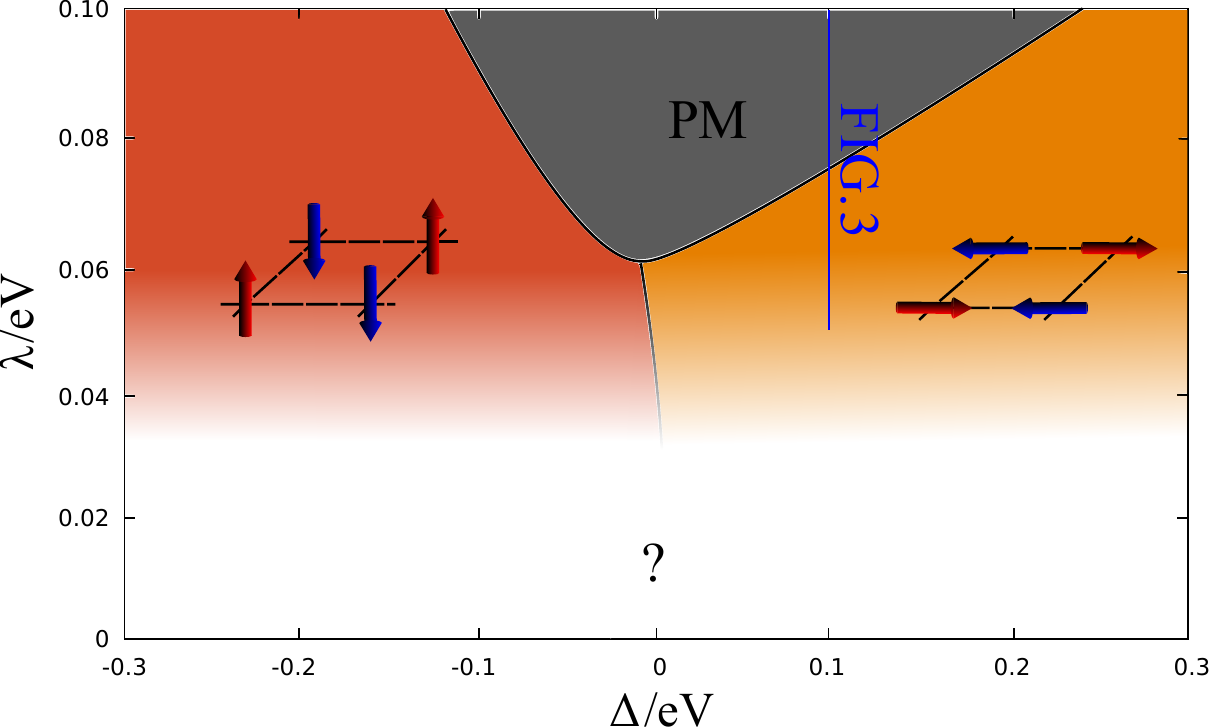}
\caption{Phase diagram for large SOC in the triplon model introduced in Sec.~\ref{sec:triplons}. The parameters are chosen to be $t_{xy}=0.2\,\text{eV},\,t_{yz}=t_{zx}=0.137\,\text{eV},\,t_{\text{NNN}}=0.1\,\text{eV}$, $U=2\,\text{eV}$, and $J_H=0.34\,\text{eV}$. For large SOC, the $J=0$ phases arises where no triplons are prevalent, while for $\Delta<0$ the z-AFM and for $\Delta>0$ the $xy$-AFM phase is favored.  \label{fig:Fig4}
}
\end{figure}
\subsubsection{$\Delta\gg\lambda$ Limit}
Presumably the most straightforward limit of the $t_{2g}^4$ model is the case
of dominant CF $\Delta \gg \lambda$ favoring the $xy$ orbital to be fully
occupied. The two remaining orbitals $zx$ and $yz$ are then half filled and
form a spin one. Magnetic ordering within the plane is then determined by the
ratio of NNN and NN superexchange, with the latter usually dominating and
favoring a checkerboard pattern.
\subsubsection{$\lambda\gg\Delta$ Limit}
For dominant SOC $\lambda\gg \Delta$, the ground
state becomes the $J=0$ state without magnetic moment and
therefore leads to a PM phase. Decreasing SOC leads to the
possibility of an admixture of the $J=1$ states to the ground state,
since the energy gap between the $J$ states is decreasing and
superexchange is driving the transition between the $J=0$ and the
$J=1$
states~\cite{PhysRevLett.111.197201,PhysRevLett.122.177201,PhysRevLett.111.197201}.
This triplon-condensation transition leads to a finite magnetization
and magnetic ordering can be described with the triplon model introduced in
Sec.~\ref{sec:triplons}.

The transition from a magnetically ordered state to the PM state is seen in
Fig.~\ref{fig:Fig3}, which shows the triplon number obtained using ED for the triplon model on a $\sqrt{8}\times\sqrt{8}$
cluster. CF is here set to $\Delta = 0.1\;\textrm{eV}$, where the magnetic
order has a checkerboard pattern.  The inflection point of the
triplon number vs. SOC $\lambda$ (at
$\lambda\approx 0.07\,$eV) was taken as the phase boundary to the PM
phase. Figure~\ref{fig:Fig3} also shows the expectation value
$\braket{J^2}$ obtained for the full spin-orbital superexchange
model. While there is no obvious signal, like, e.g., an inflection point,
for the phase boundary,  increasing $\lambda$ leads to a decrease of
$\braket{J^2}$, in agreement with the triplon number. Figure~\ref{fig:Fig4} gives the $\Delta$-$\lambda$ phase diagram for
the triplon model at intermediate to large $\lambda$, where it can be
assumed to be valid. Magnetic order switches from in-plane to
out-of-plane at $\Delta \approx 0$, and the phase diagram is in
qualitative agreement with \cite{PhysRevB.90.035137} for $\Delta \geq 0$, where  $J_H=0$ and isotropic
hopping were used. The triplon model is naturally not able to capture the physical
behavior for small SOC. From here on we will therefore focus on performing our calculations with the full spin-orbital model.

\begin{figure}
\includegraphics[width=\columnwidth]{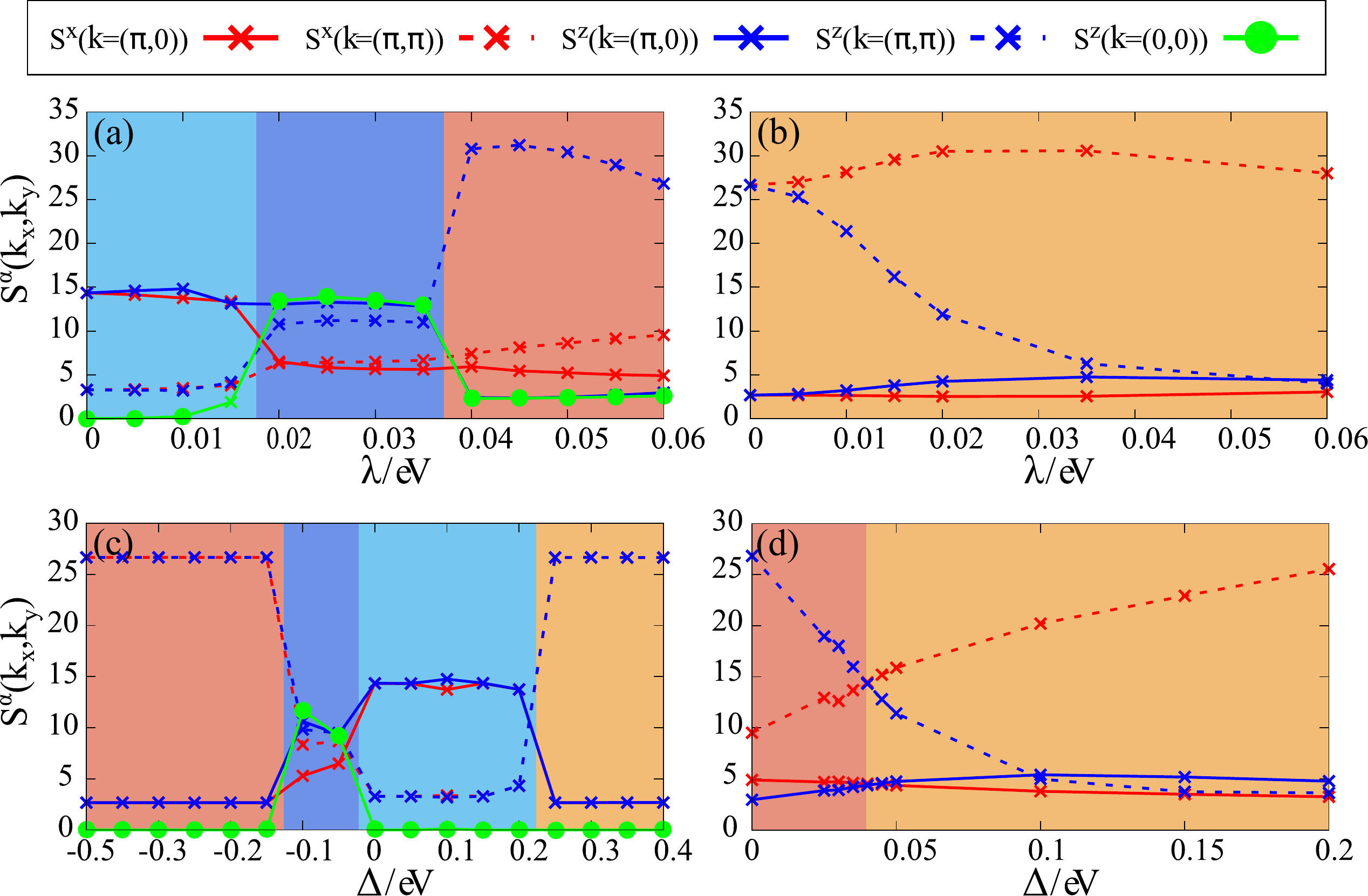}
\caption{In-plane- ($x$-$y$) and out-of-plane ($z$) SSF $\text{S}^{\alpha}(k,\lambda,\Delta)$. $\lambda$ is varied in (a) with $\Delta=0\,\text{eV}$ and
  (b) with $\Delta=0.25\,\text{eV}$.  $\Delta$ is varied in (c) with
  $\lambda=0$ and (d) with $\lambda=0.06\,\text{eV}$. The
  momenta $\mathbf{k}$ accessible on an $\sqrt{8}\times\sqrt{8}$ cluster are
  $\mathbf{k}=(0,0)$, $(\pi,0)$, $(0,\pi)$, $(\pm\tfrac{\pi}{2},\pm\tfrac{\pi}{2})$ and
  $(\pi,\pi)$. Remaining parameters are given in Sec.~\ref{sec:spinorbit}. 
  \label{fig:Fig5} 
}
\end{figure}

\subsubsection{$\lambda=0$ Limit}
 The opposite limit of $\lambda=0$ has been
investigated for varied Coulomb repulsion $U$ and Hund's coupling
$J_H$ \cite{PhysRevB.74.195124}. The calculations in \cite{PhysRevB.74.195124} were done with a
full nonperturbative treatment of the Hubbard-like Hamiltonian, which
limited the cluster size to $2\times 2$. In agreement with our
results, obtained with the full spin-orbital model, large orbital degeneracy at small CF
$0\lesssim \Delta\lesssim 0.24\;\textrm{eV}$ leads
to a complex stripy spin-orbital
pattern~\cite{PhysRevB.74.195124,PhysRevResearch.2.033201}.
For larger positive $\Delta \gtrsim 0.24\;\textrm{eV}$, CF dominates
and double occupancy is uniformly in the $xy$ orbital. Therefore the Hamiltonian reduces to
orbital-preserving terms which yield a simple Heisenberg spin Hamiltonian,
while NNN interactions are frustrated. These effects cause
a phase transition from the stripy phase to a checkerboard
pattern. 

The magnetic ordering can be inferred from the spin structure factors (SSF) for $\lambda=0$ and variable
CF that are summarized in Fig.~\ref{fig:Fig5} (c). In addition to the
stripy and $xy$-polarized checkerboard patterns seen for $\Delta
\gtrsim 0$, we find checkerboard order again for $\Delta \ll 0$. 
In this negative-$\Delta$ regime, the $xy$ orbital is half filled to
gain in-plane kinetic (resp. superexchange) energy, while double
occupation of  $xz$ and $yz$ orbitals alternate in a checkerboard
pattern as well. The overall
ordering is thus  reminiscent of that obtained for vanadates with two
$t_{2g}$ electrons~\cite{PhysRevLett.86.3879}.

\begin{figure}
\includegraphics[width=\columnwidth]{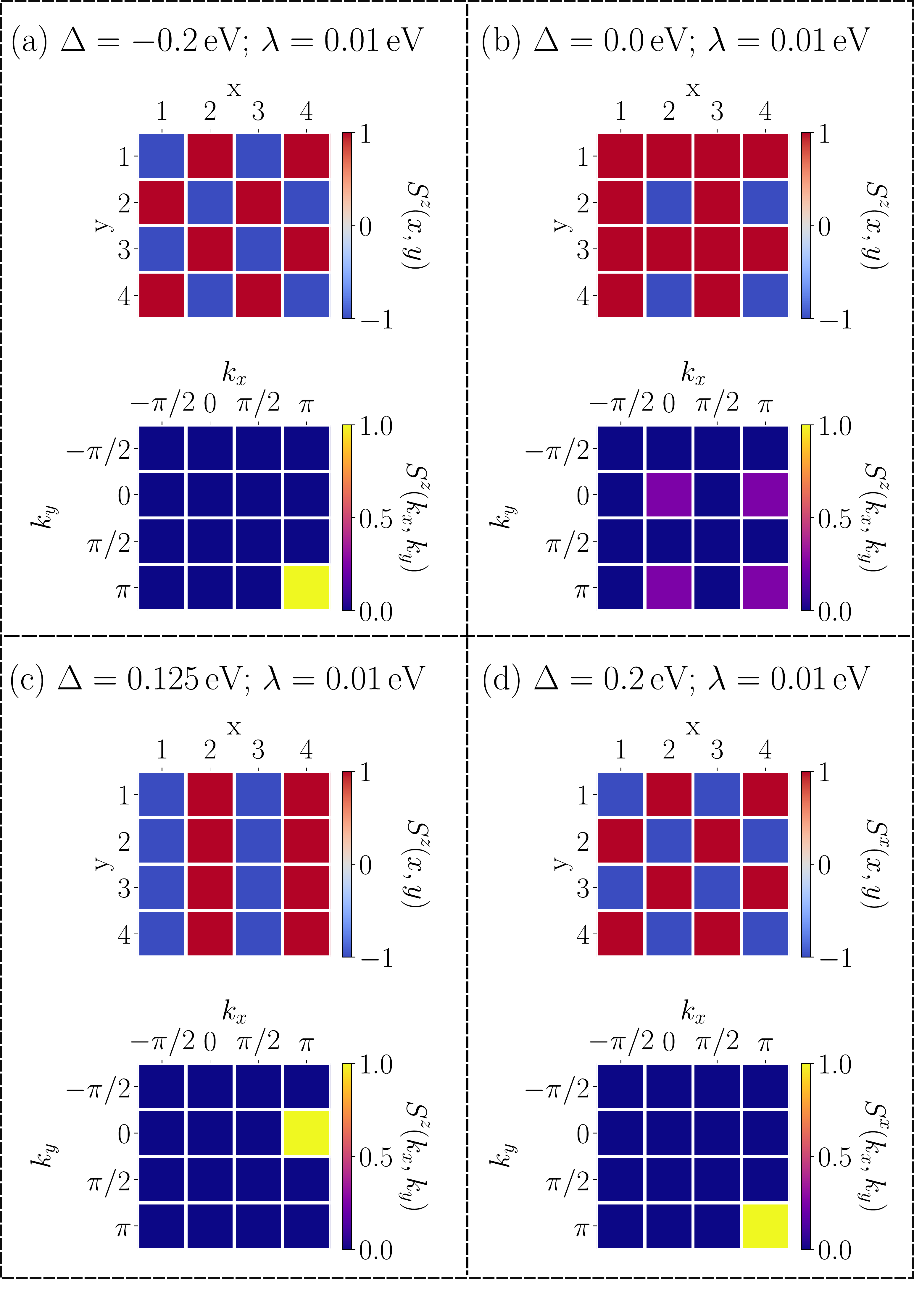}
\caption{Spin components 
  $S^z$ (a)-(c) and $S^x$ (d) per site as well as for all relevant wavevectors $\mathbf{k}$ for a 4$\times$4 square lattice. Calculations were performed with
  semiclassic parallel tempering MC. In (a)-(d) snapshots of the
  different phases arising in the parameter range $-0.2<\Delta<0.2$
  and $0.01\,\text{eV}<\lambda<0.08\,\text{eV}$ are shown. These can
  be directly compared to the ED results of
  Fig.~\ref{fig:Fig9}. \label{fig:Fig6} 
}
\end{figure}

In the regime $-0.12\;\textrm{eV}< \Delta \lesssim 0$, an
additional phase is finally seen that has finite SSF's for several momenta: 
$(\pi,0)$, $(\pi,\pi)$, and $(0,0)$. 
To clarify the nature of this phase, we performed MC
simulations on a $4\times 4$ cluster, where we include weak SOC
$\lambda=0.01\,\text{eV}$ for numerical reasons. In
Fig.~\ref{fig:Fig6} (a)-(d), snapshots of the four phases appearing in
the MC results are shown for the whole $\Delta$ range
discussed above. For $\Delta= 0$ the pattern
becomes an alternation of AFM and FM stripes [Fig.~\ref{fig:Fig6} (b)],
which leads to maxima at $S^z(\pi,0),S^z(0,\pi),S^z(\pi,\pi)$, and $
S^z(0,0)$ in the momentum space comparable to the signatures in the
SSF of the ED. This phase is from here on referred as
``3-up-1-down''.  

Overall, the  phases seen in the semiclassical model are in good
agreement with the characterization based on ED results. For $\Delta=-0.2\;\textrm{eV}$
[Fig.~\ref{fig:Fig6} (a)], the out-of-plane checkerboard AFM pattern is the ground state
with a maximum at $S^z(\pi,\pi)$ and a clear checkerboard pattern in
$z$-direction in position space. After the novel ``3-up-1-down'' phase
at $\Delta\approx 0$, positive $\Delta\approx0.125\,\text{eV}$ leads to a stripy pattern with
a maximum at $S^z(\pi,0)$ [Fig.~\ref{fig:Fig6} (c)] and larger
$\Delta=0.2\,\text{eV}$ to the  in-plane AFM order with maxima at $S^x(\pi,\pi)$
and $S^y(\pi,\pi)$, both in  accordance with the ED results. Reference~\cite{Mohapatra_2020}, which restricts itself to FM and N\'eel AFM phases, reports an FM phase with some AFM correlations at weak CF,i.e., also sees competition of FM and AFM tendencies roughly where we find the stripy and ``3-up-1-down'' patterns.

\subsection{Phase diagram of the full spin-orbital model}\label{sec:phase_diag}

After discussing the limiting cases of small and large CF and SOC, we
now  investigate the $\Delta$-$\lambda$ plane. We first study the static SSF for fixed $\lambda$ and $\Delta$
[Fig.~\ref{fig:Fig5} (a)-(d)]. Since we perform ED on an eight
site cluster, the SSF is only obtainable for eight different 
$\mathbf{k}$ values, from which only four are unique. These are
$\mathbf{k}=(0,0)$ resp. FM ordering, $(\pi,0)$ resp. stripy
ordering, $(\pi,\pi)$ resp. AFM ordering and
$(\pi/2,\pi/2)$. In Fig.~\ref{fig:Fig5} (a)-(d) only the SSF's with
appreciable weight are displayed. Our goal is an understanding of the
impact of $\lambda$ and $\Delta$ on the spin-orbital state. Hopping
parameters $t_{xy}$ , $t_{yz}$, $t_{zx}$, Coulomb repulsion $U$, and
Hund's coupling $J_{H}$ where chosen as introduced in
Sec.~\ref{sec:spinorbit}.

In Fig.~\ref{fig:Fig5} (a), CF is fixed to
$\Delta=0\,\text{eV}$ and one sees three phases depending on the
strength of SOC. For small SOC ($\lambda<0.02\,\text{eV}$), one finds
the stripy phase, where  $(\pi,0)$-SSF's have similar in-plane and out-of-plane components. This
is in concordance with the results of VCA calculations of
\cite{PhysRevResearch.2.033201} as well as ED calculations on a
$2\times 2$ cluster \cite{PhysRevB.74.195124, PhysRevLett.88.017201}.

Increasing the SOC to $0.02\,\text{eV}<\lambda<0.04\,\text{eV}$ gives
rise to a phase with contributions from in- and out-of-plane ($\pi,0$)
as well as ($\pi,\pi$) structure factors and additionally the (0,0)
out-of-plane contribution. This phase is the ``3-up-1-down'' state
already discussed in the limit $\lambda=0$ (see
Sec.~\ref{sec:LR}). Increasing SOC further ($\lambda>0.04\,\text{eV}$)
leads to an out-of-plane AFM phase. This phase is identical with the
out-of-plane AFM phase arising in the triplon model (orange phase in
Fig.~\ref{fig:Fig4}). This phase was also found at
$\Delta=0\,\text{eV}$ and substantial SOC in the VCA calculations of
\cite{PhysRevResearch.2.033201}. Further increase of SOC starts to
reduce the SSF at $(\pi,\pi)$ again, and finally suppresses all AFM
order, see the discussion of the triplon model and
Fig.~\ref{fig:Fig4}.

The results for a large fixed CF at $\Delta=0.25\,\text{eV}$ are
displayed in Fig.~\ref{fig:Fig5} (b). Starting from SOC $\lambda=0\,\text{eV}$,
the ground state is an isotropic AFM phase. SOC induces a smooth transition
to an in-plane AFM order. This is due to the fact that positive $\Delta$
favors the double occupancy of the $xy$-orbital and therefore $L^z=0$,
and as $\lambda$ couples spin and orbital momenta, this also leads to
a decrease of the $S^z$ component.

Lastly in Fig.~\ref{fig:Fig5} (d) SOC is set to the value
$\lambda=0.06\,\text{eV}$. As already mentioned earlier
$\Delta>0.04\,\text{eV}$ stabilizes an in-plane AFM pattern, due to the
preference of $L^z=0$ which results in a preference of $S^z=0$ due to
SOC. If the CF is small or has a negative sign, out-of-plane AFM ordering is
favored since the $xy$ orbital is mostly singly occupied. This
transition is well captured by the triplon model discussed in Sec.~\ref{sec:LR}. 

\begin{figure}
\includegraphics[width=\columnwidth]{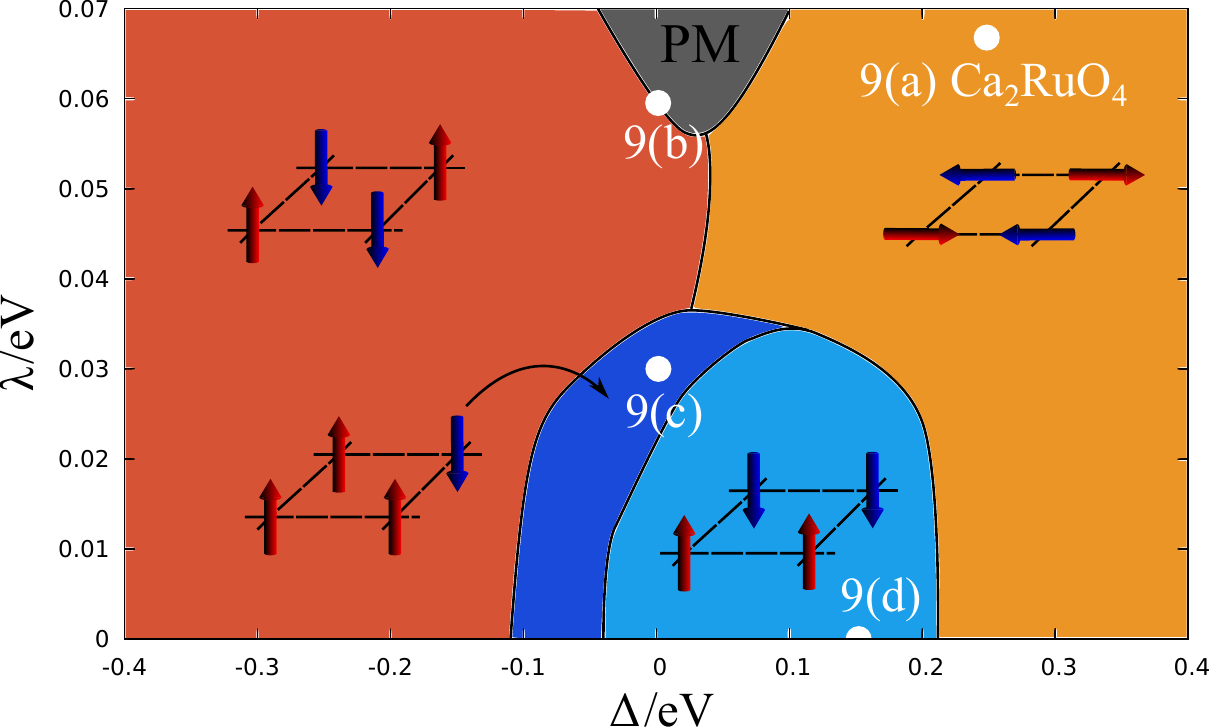}
\caption{$\Delta-\lambda$ phase diagram obtained by ED calculations on
  $\sqrt{8}\times\sqrt{8}$ cluster. The PM phase (dark grey) was
  identified via the triplon model of Sec.~\ref{sec:triplons}. Sketches
  show the spin ordering for the respective phase, calculated via MC
  on a $4\times 4$ cluster. The white dots denote the snapshots of the taken in Fig.~\ref{fig:Fig9}, to investigate the dynamical SSF (further information see Sec.~\ref{sec:Skw}), including the parameter setting for $\text{Ca}_2\text{RuO}_4$.
  calculations.   \label{fig:Fig7} 
}
\end{figure}

The information gained from the ED SSF's, supported by semiclassical MC in
case of the ``3-up-1-down'' pattern, as well as the information on the
transition to the PM phase inferred from the triplon model is
summarized in the $\Delta-\lambda$ phase diagram in Fig.~\ref{fig:Fig7}. To obtain this
phase diagram we performed multiple sweeps by varying $\Delta$ for
fixed $\lambda$ (and \textit{vice versa}), like in Fig.~\ref{fig:Fig5},
and included the PM phase from the triplon
model. We did this because the transition is better identifiable than
with $\braket{J^2}$ (see Fig.~\ref{fig:Fig3}). If both CF
and SOC are weak, the interaction terms introduced in
(\ref{eq:Eq4})-(\ref{eq:Eq6}) dominate, which favor a stripy alignment of
spins (light blue) together with a complex orbital
pattern~\cite{PhysRevB.73.094428,PhysRevResearch.2.033201}. For large CF, the double
occupation locates either at the $xy$ ($\Delta>0$) or alternates
between $zx$- and $yz$-orbitals ($\Delta<0$), which results in an
$x$-$y$-AFM (light orange) or $z$-AFM order (dark orange)
respectively.
These phases are both very robust against SOC, which favors a $J=0$ PM phase (dark grey). The competition
between the $z$-AFM and the stripy phase at small negative CF and small
SOC, causes the ``3-up-1-down'' phase to arise (dark blue).

Locating $\text{Ca}_2\text{RuO}_4$ in this phase diagram (corresponding white dot in Fig.~\ref{fig:Fig7}),
puts it solidly within the in-plane AFM phase, as expected from
experiment
\cite{PhysRevLett.115.247201,PhysRevB.100.045123,PhysRevLett.123.137204,Higgs_Ru}. Also,
$\text{Ca}_2\text{RuO}_4$ does not appear to be very close to any
CF-driven phase transition and its AFM order can thus be expected to
be rather robust.

\subsection{Comparison of semiclassical and quantum models} \label{sec:MCMC}

\begin{figure}
\includegraphics[width=\columnwidth]{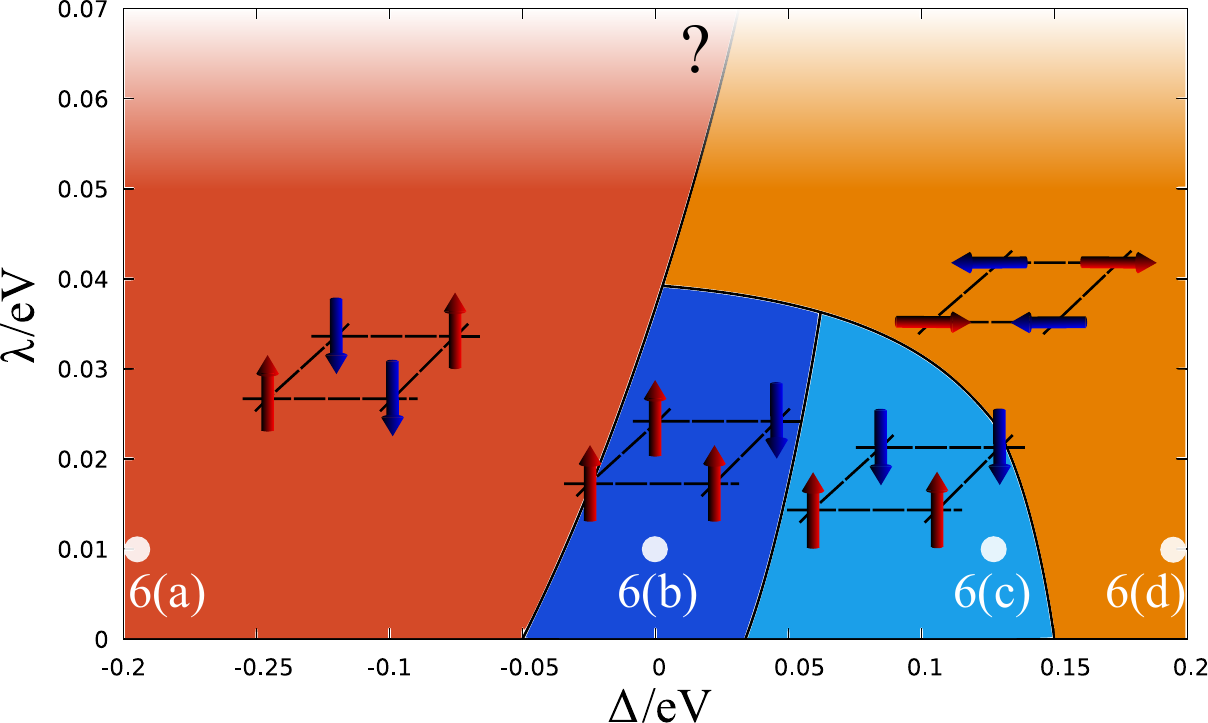}
\caption{Phase diagram depending on $\lambda$ and $\Delta$ obtained
  via MC for remaining parameters as given in Sec.~\ref{sec:spinorbit}. 
  White dots denote the
  snapshots of Fig.~\ref{fig:Fig6}.\label{fig:Fig8}}
\end{figure}

Having made use of a semiclassical Markov chain MC to identify the ``3-up-1-down'' phase, we now compare the
semiclassical and quantum phase diagrams more generally. Several snapshots
for weak SOC were already discussed in Sec.~\ref{sec:LR}  to clarify
the regime of weak SOC and $\Delta\lesssim 0$. To obtain a full
semiclassical phase diagram, several CF sweeps for different strengths of
SOC between $-0.2\,\text{eV}<\Delta<0.2\,\text{eV}$ were performed and
the SSF for $(\pi,\pi)$, $(\pi,0)$ and $(0,0)$ are calculated. The
results yield the phase diagram shown in Fig.~\ref{fig:Fig8} [white
  points denote the locations of the snapshots of Fig.~\ref{fig:Fig6}
  (a)-(d)]. For dominant SOC $\lambda>0.04\,\text{eV}$ and $\Delta<0$, 
an out-of-plane AFM phase arises (dark orange in Fig.~\ref{fig:Fig8}),
while positive CF $\Delta>0\,\text{eV}$ gives rise to an in-plane AFM phase
(light orange). For strong CF $|\Delta|>0$ both phases stay stable up
to $\lambda=0$. For weak SOC $\lambda<0.04\,\text{eV}$ and CF
$0.05\,\text{eV}<\Delta<0.15\,\text{eV}$, the interaction part in the
Hamiltonian becomes dominant. This is similar to Sec.~\ref{sec:phase_diag}, which
leads to an out-of-plane stripy phase (light blue). The competition
between the stripy and the out-of-plane AFM phase leads to the ``3-up-1-down''
phase (dark blue) already discussed in Sec.~\ref{sec:LR} at
$-0.05\,\text{eV}<\Delta<0.05\,\text{eV}$ for
$\lambda<0.04\,\text{eV}$.

This phase diagram is in good 
qualitative agreement with the spin-orbit model
(Fig.~\ref{fig:Fig4}). The exact location of the phase transitions
differ somewhat between semiclassical and quantum
models. In comparison to the ED simulations, in the
semiclassical calculations the AFM phases (both in- and out-of-plane) are
more dominant. While ED predicts the $z$-AFM phase to end at
$\Delta\approx -0.1\,\text{eV}$ for $\lambda=0\,\text{eV}$, in the semiclassical
simulations the $z$-AFM phase stays robust until $\Delta\approx
-0.05\,\text{eV}$ (same for the $x-y$-AFM ordering see Fig.~\ref{fig:Fig7} and
Fig.~\ref{fig:Fig8}).
While the origin for the the difference might lie in the small
clusters used (especially for ED), it is quite plausible that quantum
fluctuations have the strongest impact near orbital degeneracy.
The fact that semiclassical MC captures the
same phases as ED, gives a promising pathway that effective
spin-orbital models can also be studied on significant larger cluster
size with semiclassical MC while still giving reasonable results.

\subsection{Dynamic spin-structure factor} \label{sec:Skw}
\begin{figure}
\includegraphics[width=\columnwidth]{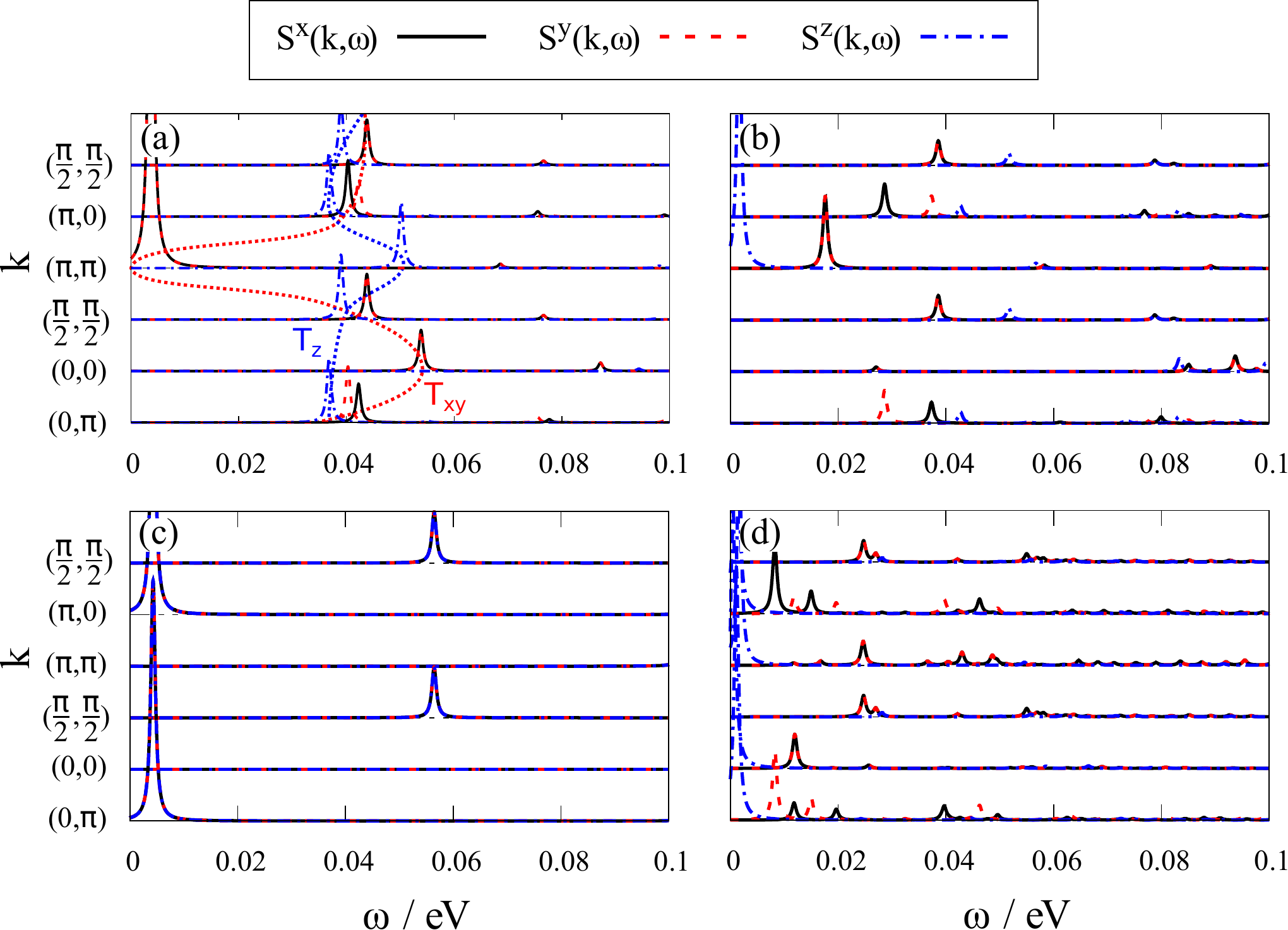}
\caption{Dynamical spin structure factor $S(\mathbf{k},\omega)$ for
  (a) $\Delta=0.25\,\text{eV};\lambda=0.065\,\text{eV}$,
  (b) $\Delta=0.0\,\text{eV};\lambda=0.06\,\text{eV}$, (c)
  $\Delta=0.15\,\text{eV};\lambda=0\,\text{eV}$, and (d)
  $\Delta=0.0\,\text{eV};\lambda=0.03\,\text{eV}$. Parameters in (a)
  are the ones used to describe  $\text{Ca}_2\text{RuO}_4$~\cite{PhysRevResearch.2.033201}
  and   capture the characteristics of INS experiments. \label{fig:Fig9} 
}
\end{figure}

Experimentally, the various phases might be distinguished via magnetic
excitations. Therefore we discuss here the signatures expected for the
dynamic spin-structure factor 
\begin{align}\label{eq:Eq12}
S^{\alpha}(\mathbf{k},\omega)=-\frac{1}{\pi}\text{Im}\bra{\phi_0}S^{\alpha}(-\mathbf{k})\frac{1}{\omega-H+i0^+}S^{\alpha}(\mathbf{k})\ket{\phi_0},
\end{align}
which gives an $\omega$ resolution of the phases introduced in
Fig.~\ref{fig:Fig5}. This can then be compared to inelastic neutron scattering
\cite{PhysRevLett.115.247201,Higgs_Ru}. 
In Fig.~\ref{fig:Fig9} the DSSF's of
the four distinct phases are shown. The locations of these snapshots in the phase diagram are denoted with white dots in Fig.\ref{fig:Fig7}.

\subsubsection{Excitations of the in-plane AFM regime} \label{sec:Skw_xy}

For
$\Delta=0.25\,\text{eV}$ and $\lambda=0.065\,\text{eV}$ [Fig.~\ref{fig:Fig9}
  (a)], the the Goldstone mode at
$(\pi,\pi)$ allows us to identify the in-plane AFM phase found above in
Fig.~\ref{fig:Fig5} (b) and (d). The spectrum of Fig.~\ref{fig:Fig9}(a)
was already presented in Ref.~\cite{PhysRevResearch.2.033201} as the
parameters closely fit 
Ca\textsubscript{2}RuO\textsubscript{4}. As already discussed in \cite{PhysRevResearch.2.033201} the in-plane
(red guideline) and out-of-plane (blue guideline) transverse  modes
can be identified. Especially the in-plane transverse mode shows an
excellent agreement to \cite{Higgs_Ru} reproducing the maximum at
$\mathbf{k}=(0,0)$ and $\omega=0.54\,\text{eV}$.

\begin{figure}
\includegraphics[width=\columnwidth]{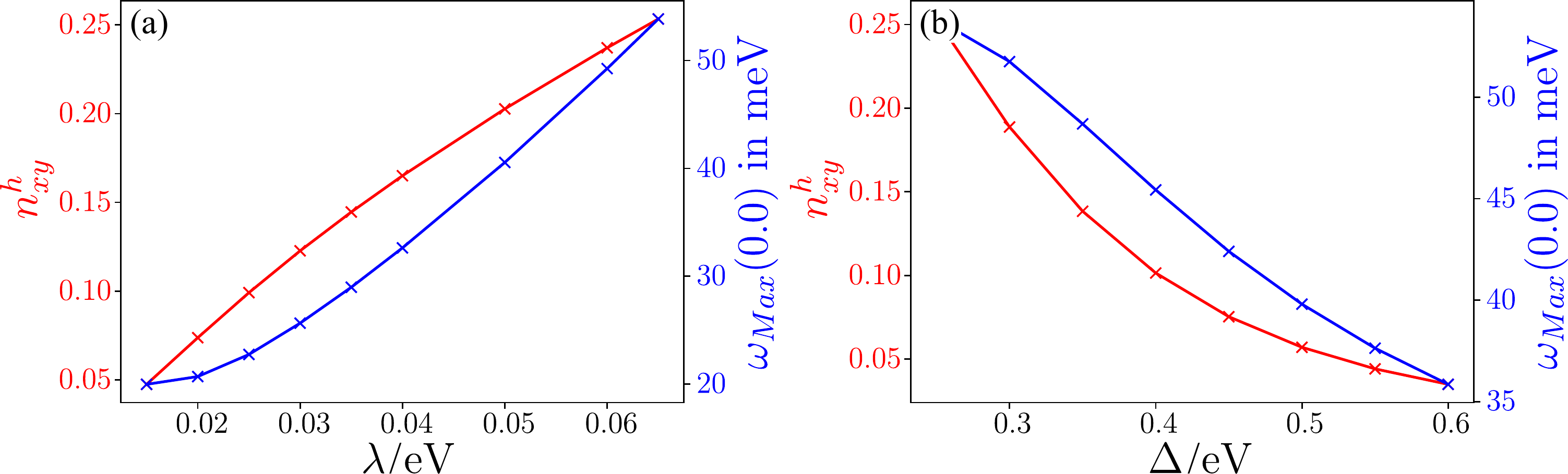}
\caption{Excitation energy $\omega_{\textrm{max}}$ at
  $\mathbf{k}=(0,0)$ (blue) and hole density $n^h_{xy}$ in the $xy$
  orbital (red). (a) Depending on SOC for
  CF $\Delta=0.25\,\text{eV}$ and (b) depending on CF for SOC
  $\lambda=0.065\,\text{eV}$. \label{fig:Fig10}}
\end{figure}

This maximum, a characteristic
signature of the $x$-$y$ symmetry of the  magnetic moments, strongly depends on the hole density $n^h_{xy}$
in the $xy$-orbital, which is $n^h_{xy}\approx 0.25$ in
Fig.~\ref{fig:Fig9}(a). Figure~\ref{fig:Fig10}(a) shows the
dependence of $n^h_{xy}$ and of the excitation energy
$\omega_{\text{Max}}(0,0)$ on SOC $\lambda$. The excitation energy at
$\mathbf{k}=(0,0)$ increases steadily from a minimum at
$\omega\approx0.02\,\text{eV}$ to the maximum $\omega=0.54\,\text{eV}$
for the $\text{Ca}_2\text{RuO}_4$ parameters in Fig.~\ref{fig:Fig9}
(a). Having a  maximum at
$\mathbf{k}=(0,0)$ is thus closely connected to finite - but not
necessarily large - hole density in the $xy$-orbital. 

Without SOC, strong CF
$\Delta=0.25\,\text{eV}$  localizes the two holes in the $zx$- and $yz$-orbital,
with $n_{xy}^{h}\approx 0.05$, in agreement with ab-initio
calculations for Ca$_2$RuO$_4$ performed without SOC~\cite{PhysRevB.95.075145,hund_Ca2RuO4}. Increasing SOC softens
this polarization because it couples $\mathbf{S}$ and
$\mathbf{L}$ and thus competes with $\Delta$. SOC increases
the hole density at $xy$ so that it reaches $n_{xy}^h=0.25$ at
$\lambda=0.065\,$eV. On one hand, this implies that the $xy$-orbital
continues to be rather close to fully occupied and justifies the
picture of Ca$_2$RuO$_4$ as orbitally
ordered~\cite{PhysRevB.101.205128}. On the other hand,
Figs.~\ref{fig:Fig10}(a) and~\ref{fig:Fig9}(a) reveal that the
relatively few holes in the $xy$-orbital have a decisive impact on 
magnetic excitations. 

\textit{Vice versa}, if SOC is fixed and the CF is increased
[Fig.~\ref{fig:Fig10} (b)] the maximum at $\mathbf{k}=(0,0)$
vanishes. Starting at $\lambda=0.065\,$eV and $\Delta=0.25\,$eV
the maximum is, as already discussed, at $\omega=54\,$meV.
Increasing $\Delta$ up to $\Delta=0.6\,$eV strongly suppresses
the hole density in the $xy$-orbital and at the same time leads to a minimum
in the excitation spectrum  at $\mathbf{k}=(0,0)$ and
$\omega=36\,$meV. It is of note that while the hole density
appears to be linked to $\omega$ at $\mathbf{k}=(0,0)$, it is
not the only influence. This can be concluded by the fact that
for the parameter settings
$\Delta=0.25\,\text{eV};\,\lambda=0.015\,\text{eV}$ and
$\Delta=0.6\,\text{eV};\,\lambda=0.065\,\text{eV}$ the hole
densities are very similar ($n_{xy}^h\approx 0.05$) while
$\omega_{\text{Max}}(0,0)$ of the excitation differs by a
factor of $1.8$ between strong and weak values of SOC and
CF. This means that SOC and CF also have direct influence to
the excitation at $\mathbf{k}=(0,0)$ in addition to the indirect
influence via the hole density of $n_{xy}^{h}$.

Taken together, the extensive study of the excitation at
$\mathbf{k}=(0,0)$ has shown that excitation
spectra already differ from the one measured in
\cite{Higgs_Ru} for relatively weak changes in $\lambda$ and
$\Delta$, even though  the
ground state of $\text{Ca}_2\text{RuO}_4$ is quite robust
against such perturbations.  It is therefore remarkable that the DSSF in
Fig.~\ref{fig:Fig9}(a) of the effective model is in such close
agreement with the experimental data.  

\subsubsection{Excitations of the PM and various out-of-plane AFM phases} \label{sec:Skw_z}

\begin{figure}
\includegraphics[width=\columnwidth]{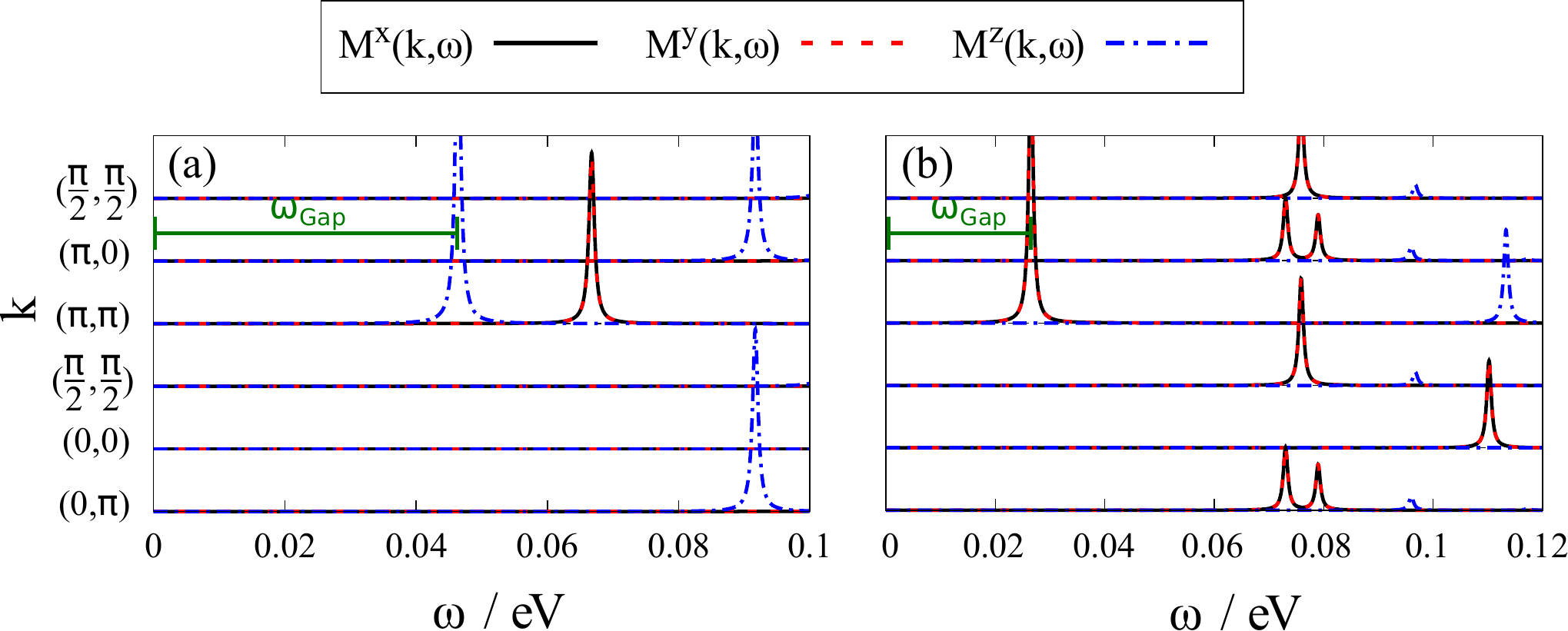}
\caption{Dynamical magnetic structure factor
  $\mathbf{M}(\mathbf{k},\omega)$ for  (a) $\Delta=0.0\,$eV and (b)
  $\Delta=0.25\,$eV with substantial SOC
  $\lambda=0.12\,$eV. $\omega_{\text{Gap}}$ marks the energy gap between the
  ground state and the lowest lying
  excitation.\label{fig:Fig11} 
}
\end{figure}

Decreasing CF to $\Delta=0.0\,\text{eV}$ and leaving
$\lambda=0.06\,\text{eV}$, the lowest excitation only has out-of-plane
contributions [Fig.~\ref{fig:Fig9}(b)]. This indicates $z$-AFM ordering, cf. Fig.~\ref{fig:Fig5} (a) and
(d), although the system is here close to the PM state, see Fig.~\ref{fig:Fig7}.
Choosing a large value for SOC $\lambda=0.12\,\text{eV}$ firmly puts
the system into the PM state, and the
excitation minimum at $(\pi,\pi)$ moves to higher $\omega$. This can be seen in
Fig.~\ref{fig:Fig11}(a), with the magnetization $\mathbf{M}=2\mathbf{S}-\mathbf{L}$ and the dynamical magnetic structure factor obtained analogue to (\ref{eq:Eq12}). The excitation gap is
$\omega_{\text{Gap}}=0.046\,\text{eV}$ [Fig.~\ref{fig:Fig11}(a)]
meaning there is a significant energy cost for the system to create a
triplon.

Increasing the CF to $\Delta=0.25\,\text{eV}$ [Fig.~\ref{fig:Fig11}
  (b)] one can see that (i) the lowest-energy triplon has now $x$-$y$
character and (ii) its energy is decreased
significantly to $\omega_{\text{Gap}}=0.027\,\text{eV}$. The finite CF
 thus reduces triplon energy so that they can eventually
condense into magnetic order. This can also be seen nicely in Fig.~\ref{fig:Fig7} where
$\text{Ca}_2\text{RuO}_4$ (corresponding white dot in Fig.~\ref{fig:Fig7}) would be in the PM phase if it had
no significant CF splitting. 

Spectra for the stripy and and  ``3-up-1-down'' phases realized near
orbital degeneracy are shown in Fig.~\ref{fig:Fig9} (c) resp. (d). The
stripy phase ($\Delta=0.15\,\text{eV}; \lambda=0\,\text{eV}$)
in Fig.~\ref{fig:Fig9} (c) not only shows spin isotropy but also a
degeneracy between $x$- ($\pi,0$) and $y$-stripy ($0,\pi$)
order. Finally, the DSSF of the ``3-up-1-down''
phase from Fig.~\ref{fig:Fig5} (a) is displayed in Fig.~\ref{fig:Fig9}
(d) and shows the many ordering vectors contributing for $\omega \to 0$.  

\begin{figure}
  \includegraphics[width=\columnwidth]{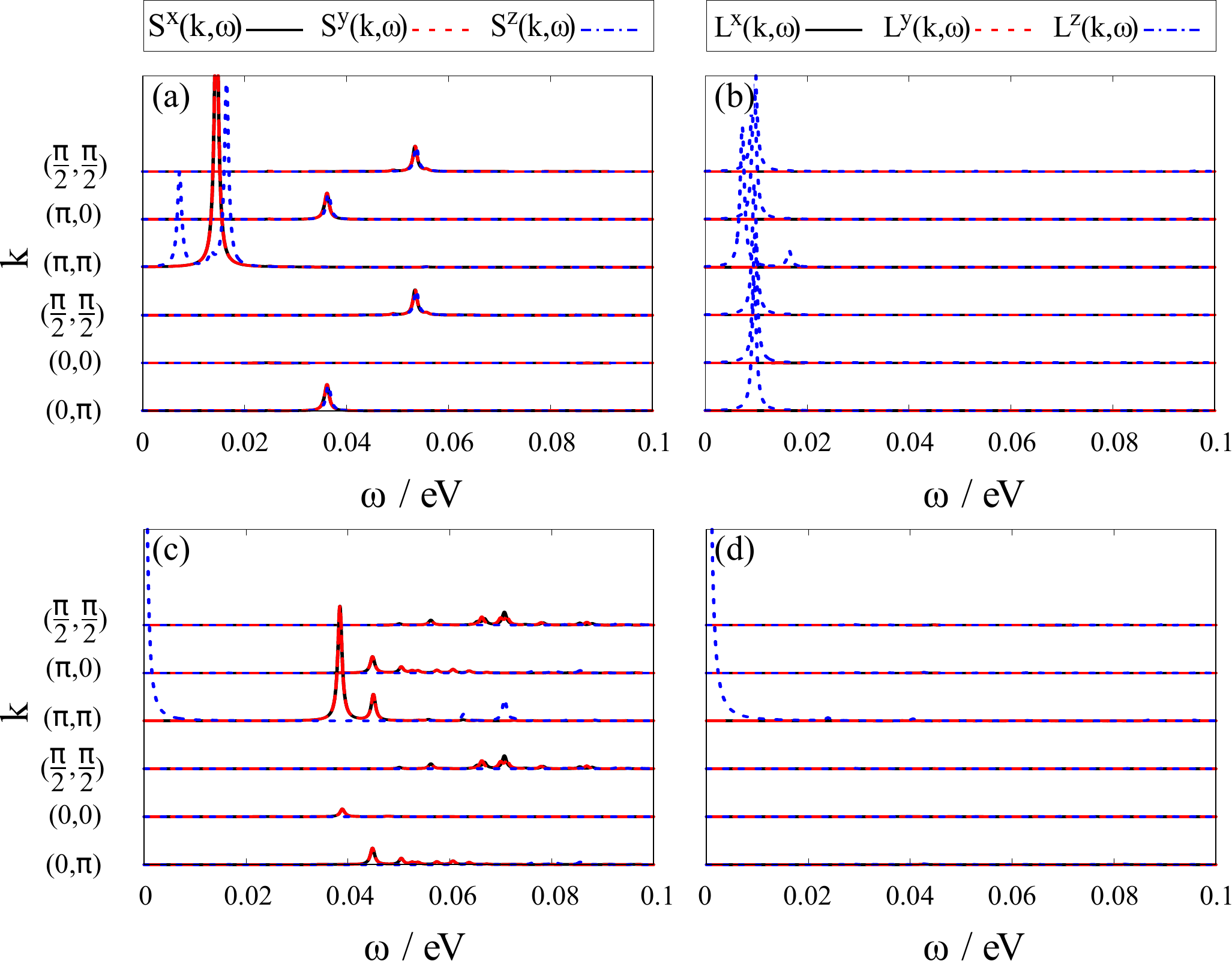}
  \caption{Dynamical Spin and Orbital structure factors for crystal
    field $\Delta = -0.3\;\textrm{eV}$ and weak SOC. (a) and (c) show
    the DSSF (\ref{eq:Eq12}) while (b) and (d) give the orbital
    analogue based on (\ref{eq:Eq3}). In (a) and (b),
    $\lambda=0.002\;\textrm{eV}$ and in (c) and (d),
    $\lambda=0.01\;\textrm{eV}$.\label{fig:Fig12}
    }
\end{figure}

The last phase to be discussed in detail is the checkerboard AFM order
with out-of-plane anisotropy at $\Delta \lesssim 0$. For $\lambda=0$,
moderate CF $\Delta \approx -0.3\;\textrm{eV}$ is enough to fix the
$xy$ orbital to half filling, so that either $xz$ or $yz$
orbitals are double occupied. These two states alternate in a
checkerboard pattern with the same unit cell as a Heisenberg-symmetric
AFM. A corresponding magnetic excitation spectrum is shown in
Fig.~\ref{fig:Fig12}(a), where weak $\lambda =
0.0002\;\textrm{eV}$ induces slight Ising anisotropy into a nearly
isotropic spectrum. For the orbital analogue to the DSSF, the spin
operator $S^{\alpha}$ in (\ref{eq:Eq12}) is replaced by
angular-momentum operators (\ref{eq:Eq3}). The resulting spectrum shown
in  Fig.~\ref{fig:Fig12}(b) is, however, featureless, because
alternating order in real orbitals is \emph{quadrupolar} and would
show up in the $(L^x)^2-(L^y)^2\propto n_{xz}-n_{yz}$ channel. 

Already for rather small $\lambda = 0.01\;\textrm{eV}$, however, 
Ising anisotropy in spin excitations is very pronounced 
with an ordered moment along $z$ and a substantial excitation gap,
see Fig.~\ref{fig:Fig12}(c). At the same time, orbital order is
now also clearly dipolar and peaked at $(\pi,\pi)$, see
Fig.~\ref{fig:Fig12}(d). SOC has thus coupled spin and orbital
ordering into a checkerboard pattern with $L^z=1$, $S^z=-1$ in one
sublattice and $L^z=-1$, $S^z=1$ on the other.  In contrast to $\Delta > 0$ discussed above, where SOC induces a
gradual crossover from a Heisenberg spin-one system to an excitonic AFM state,
the transition between the isotropic and Ising states is here much
more abrupt. 

\section{Discussion and Conclusions}\label{sec:conclusions} 

In this paper we investigate an effective low-energy spin-orbital
Hamiltonian for spin-orbit coupled Mott insulators like
$\text{Ca}_2\text{RuO}_4$. This model interpolates from the strong-SOC
regime, where a description in terms of triplons is applicable, to
vanishing SOC and moreover includes Hund's coupling and anisotropic
hopping. For this model, we performed ED calculations on 
a $\sqrt{8}\times\sqrt{8}$ square lattice to obtain both static and
dynamic SSF's for varying CF $\Delta$ and SOC
$\lambda$. The results for the static SSF indicated the existence of four
distinct phases. Namely a $z$-AFM and $xy$-AFM with checkerboard
pattern, stripy-AFM  and a
``3-up-1-down'' phase at small CF and SOC $\lambda\gtrsim 0$. The stripy and
``3-up-1-down'' arise near orbital degeneracy, i.e., when neither SOC
nor CF dominate, out of the competition and partial frustration of
various superexchange terms. The 
two checkerboard phases, in contrast, extend to large CF's and include
excitonic variants at moderate SOC, whereas strong SOC finally drives a transition
to a PM state. 

We supplemented the ED analysis of the full quantum model with  MC
calculations for a semiclassical variant of the same spin-orbital model on
a $4\times 4$ cluster. Overall agreement between the semiclassical and
quantum-mechanical models was quite good, with the largest differences
found around orbital degeneracy $\Delta, \lambda \approx 0$. 
The transition to the PM $J=0$ phase at strong
SOC coupling was clarified  with the help of an effective triplon model
comparable to \cite{PhysRevLett.111.197201}. Combining these results
gave us a complete $\Delta-\lambda$ phase diagram that establishes the
competition of CF and SOC for strongly correlated $t_{2g}^4$ systems.

We also investigate the DSSF and
show that there is a remarkable correspondence~\cite{PhysRevResearch.2.033201} between calculations based on
\textit{ab initio} parameters and neutron-scattering results, despite the fact that the calculations appear to
strongly depend on the hole density in the $xy$-orbital. Parameter
dependence is also quite sensitive, which makes this a stringent test
of the model that allows a distinction between orbital degeneracy
lifted by a CF or by SOC. We further give
spectra expected for the other phases found with the model.

In
contrast to the gradual impact of SOC on the excitations  of the
orbitally polarized regime $\Delta>0$, a much clearer transition is
revealed at $\Delta<0$. Relatively small SOC is enough to switch
from alternating orbital order and Heisenberg AFM to order involving
complex orbitals. However, coupling to further lattice distortions,
not discussed here, would be expected to push this transition to
stronger SOC. This physics might also be relevant to $t_{2g}^2$
systems, i.e.,  with two electrons as in vanadates, where similar
alternating or orbital order
arises~\cite{PhysRevLett.86.3879}. Although SOC for two electrons has opposite sign
than the two-hole case discussed here, this does not qualitatively
affect results in the parameter regime with effective Ising
symmetry, i.e. when  $\Delta<0$  leads to nearly  empty (for $t_{2g}^2$) resp. always doubly
occupied (for $t_{2g}^4$) $xy$ orbitals. The spin-orbital
superexchange model discussed here can naturally be extended to the
two-electron case also beyond this regime.

\begin{acknowledgments}
The authors acknowledge support by the state of Baden-W\"urttemberg through
bwHPC and via the Center for Integrated Quantum Science and Technology
(IQST). M.D. thanks KITP at UCSB for kind hospitality, this research was thus
supported in part by the National Science Foundation under Grant No. NSF
PHY-1748958. 
\end{acknowledgments}

\bibliography{References}

\begin{thebibliography}{44}%
\makeatletter
\providecommand \@ifxundefined [1]{%
 \@ifx{#1\undefined}
}%
\providecommand \@ifnum [1]{%
 \ifnum #1\expandafter \@firstoftwo
 \else \expandafter \@secondoftwo
 \fi
}%
\providecommand \@ifx [1]{%
 \ifx #1\expandafter \@firstoftwo
 \else \expandafter \@secondoftwo
 \fi
}%
\providecommand \natexlab [1]{#1}%
\providecommand \enquote  [1]{``#1''}%
\providecommand \bibnamefont  [1]{#1}%
\providecommand \bibfnamefont [1]{#1}%
\providecommand \citenamefont [1]{#1}%
\providecommand \href@noop [0]{\@secondoftwo}%
\providecommand \href [0]{\begingroup \@sanitize@url \@href}%
\providecommand \@href[1]{\@@startlink{#1}\@@href}%
\providecommand \@@href[1]{\endgroup#1\@@endlink}%
\providecommand \@sanitize@url [0]{\catcode `\\12\catcode `\$12\catcode
  `\&12\catcode `\#12\catcode `\^12\catcode `\_12\catcode `\%12\relax}%
\providecommand \@@startlink[1]{}%
\providecommand \@@endlink[0]{}%
\providecommand \url  [0]{\begingroup\@sanitize@url \@url }%
\providecommand \@url [1]{\endgroup\@href {#1}{\urlprefix }}%
\providecommand \urlprefix  [0]{URL }%
\providecommand \Eprint [0]{\href }%
\providecommand \doibase [0]{https://doi.org/}%
\providecommand \selectlanguage [0]{\@gobble}%
\providecommand \bibinfo  [0]{\@secondoftwo}%
\providecommand \bibfield  [0]{\@secondoftwo}%
\providecommand \translation [1]{[#1]}%
\providecommand \BibitemOpen [0]{}%
\providecommand \bibitemStop [0]{}%
\providecommand \bibitemNoStop [0]{.\EOS\space}%
\providecommand \EOS [0]{\spacefactor3000\relax}%
\providecommand \BibitemShut  [1]{\csname bibitem#1\endcsname}%
\let\auto@bib@innerbib\@empty
\bibitem [{\citenamefont {Witczak-Krempa}\ \emph {et~al.}(2014)\citenamefont
  {Witczak-Krempa}, \citenamefont {Chen}, \citenamefont {Kim},\ and\
  \citenamefont {Balents}}]{doi:10.1146/annurev-conmatphys-020911-125138}%
  \BibitemOpen
  \bibfield  {author} {\bibinfo {author} {\bibfnamefont {W.}~\bibnamefont
  {Witczak-Krempa}}, \bibinfo {author} {\bibfnamefont {G.}~\bibnamefont
  {Chen}}, \bibinfo {author} {\bibfnamefont {Y.~B.}\ \bibnamefont {Kim}},\ and\
  \bibinfo {author} {\bibfnamefont {L.}~\bibnamefont {Balents}},\ }\bibfield
  {title} {\bibinfo {title} {Correlated quantum phenomena in the strong
  spin-orbit regime},\ }\href
  {https://doi.org/10.1146/annurev-conmatphys-020911-125138} {\bibfield
  {journal} {\bibinfo  {journal} {Annual Review of Condensed Matter Physics}\
  }\textbf {\bibinfo {volume} {5}},\ \bibinfo {pages} {57} (\bibinfo {year}
  {2014})},\ \Eprint
  {https://arxiv.org/abs/https://doi.org/10.1146/annurev-conmatphys-020911-125138}
  {https://doi.org/10.1146/annurev-conmatphys-020911-125138} \BibitemShut
  {NoStop}%
\bibitem [{\citenamefont {Rau}\ \emph {et~al.}(2016)\citenamefont {Rau},
  \citenamefont {Lee},\ and\ \citenamefont
  {Kee}}]{doi:10.1146/annurev-conmatphys-031115-011319}%
  \BibitemOpen
  \bibfield  {author} {\bibinfo {author} {\bibfnamefont {J.~G.}\ \bibnamefont
  {Rau}}, \bibinfo {author} {\bibfnamefont {E.~K.-H.}\ \bibnamefont {Lee}},\
  and\ \bibinfo {author} {\bibfnamefont {H.-Y.}\ \bibnamefont {Kee}},\
  }\bibfield  {title} {\bibinfo {title} {Spin-orbit physics giving rise to
  novel phases in correlated systems: Iridates and related materials},\ }\href
  {https://doi.org/10.1146/annurev-conmatphys-031115-011319} {\bibfield
  {journal} {\bibinfo  {journal} {Annual Review of Condensed Matter Physics}\
  }\textbf {\bibinfo {volume} {7}},\ \bibinfo {pages} {195} (\bibinfo {year}
  {2016})},\ \Eprint
  {https://arxiv.org/abs/https://doi.org/10.1146/annurev-conmatphys-031115-011319}
  {https://doi.org/10.1146/annurev-conmatphys-031115-011319} \BibitemShut
  {NoStop}%
\bibitem [{\citenamefont {Pesin}\ and\ \citenamefont
  {Balents}(2010)}]{Pesin:2010ju}%
  \BibitemOpen
  \bibfield  {author} {\bibinfo {author} {\bibfnamefont {D.}~\bibnamefont
  {Pesin}}\ and\ \bibinfo {author} {\bibfnamefont {L.}~\bibnamefont
  {Balents}},\ }\bibfield  {title} {\bibinfo {title} {{{Mott physics and band
  topology in materials with strong spin-orbit interaction}}},\ }\href
  {https://doi.org/doi:10.1038/nphys1606} {\bibfield  {journal} {\bibinfo
  {journal} {Nature Physics}\ }\textbf {\bibinfo {volume} {6}},\ \bibinfo
  {pages} {376} (\bibinfo {year} {2010})}\BibitemShut {NoStop}%
\bibitem [{\citenamefont {Chaloupka}\ \emph
  {et~al.}(2010{\natexlab{a}})\citenamefont {Chaloupka}, \citenamefont
  {Jackeli},\ and\ \citenamefont {Khaliullin}}]{Chaloupka:2010gi}%
  \BibitemOpen
  \bibfield  {author} {\bibinfo {author} {\bibfnamefont {J.}~\bibnamefont
  {Chaloupka}}, \bibinfo {author} {\bibfnamefont {G.}~\bibnamefont {Jackeli}},\
  and\ \bibinfo {author} {\bibfnamefont {G.}~\bibnamefont {Khaliullin}},\
  }\bibfield  {title} {\bibinfo {title} {Kitaev-heisenberg model on a honeycomb
  lattice: Possible exotic phases in iridium oxides
  {${\mathrm{A}}_{2}{\mathrm{IrO}}_{3}$}},\ }\href
  {https://doi.org/10.1103/PhysRevLett.105.027204} {\bibfield  {journal}
  {\bibinfo  {journal} {Phys. Rev. Lett.}\ }\textbf {\bibinfo {volume} {105}},\
  \bibinfo {pages} {27204} (\bibinfo {year} {2010}{\natexlab{a}})}\BibitemShut
  {NoStop}%
\bibitem [{\citenamefont {Kitaev}(2006)}]{Kitaev:2006ik}%
  \BibitemOpen
  \bibfield  {author} {\bibinfo {author} {\bibfnamefont {A.}~\bibnamefont
  {Kitaev}},\ }\bibfield  {title} {\bibinfo {title} {{{Anyons in an exactly
  solved model and beyond}}},\ }\href
  {https://doi.org/10.1016/j.aop.2005.10.005} {\bibfield  {journal} {\bibinfo
  {journal} {Annals of Physics}\ }\textbf {\bibinfo {volume} {321}},\ \bibinfo
  {pages} {2} (\bibinfo {year} {2006})}\BibitemShut {NoStop}%
\bibitem [{\citenamefont {Bertinshaw}\ \emph
  {et~al.}(2019{\natexlab{a}})\citenamefont {Bertinshaw}, \citenamefont {Kim},
  \citenamefont {Khaliullin},\ and\ \citenamefont {Kim}}]{rev_square_iri}%
  \BibitemOpen
  \bibfield  {author} {\bibinfo {author} {\bibfnamefont {J.}~\bibnamefont
  {Bertinshaw}}, \bibinfo {author} {\bibfnamefont {Y.}~\bibnamefont {Kim}},
  \bibinfo {author} {\bibfnamefont {G.}~\bibnamefont {Khaliullin}},\ and\
  \bibinfo {author} {\bibfnamefont {B.}~\bibnamefont {Kim}},\ }\bibfield
  {title} {\bibinfo {title} {Square lattice iridates},\ }\href
  {https://doi.org/10.1146/annurev-conmatphys-031218-013113} {\bibfield
  {journal} {\bibinfo  {journal} {Annu. Rev. Condens. Matter Phys.}\ }\textbf
  {\bibinfo {volume} {10}},\ \bibinfo {pages} {315} (\bibinfo {year}
  {2019}{\natexlab{a}})}\BibitemShut {NoStop}%
\bibitem [{\citenamefont {Winter}\ \emph {et~al.}(2017)\citenamefont {Winter},
  \citenamefont {Tsirlin}, \citenamefont {Daghofer}, \citenamefont {van~den
  Brink}, \citenamefont {Singh}, \citenamefont {Gegenwart},\ and\ \citenamefont
  {Valenti}}]{0953-8984-29-49-493002}%
  \BibitemOpen
  \bibfield  {author} {\bibinfo {author} {\bibfnamefont {S.~M.}\ \bibnamefont
  {Winter}}, \bibinfo {author} {\bibfnamefont {A.~A.}\ \bibnamefont {Tsirlin}},
  \bibinfo {author} {\bibfnamefont {M.}~\bibnamefont {Daghofer}}, \bibinfo
  {author} {\bibfnamefont {J.}~\bibnamefont {van~den Brink}}, \bibinfo {author}
  {\bibfnamefont {Y.}~\bibnamefont {Singh}}, \bibinfo {author} {\bibfnamefont
  {P.}~\bibnamefont {Gegenwart}},\ and\ \bibinfo {author} {\bibfnamefont
  {R.}~\bibnamefont {Valenti}},\ }\bibfield  {title} {\bibinfo {title} {{Models
  and materials for generalized Kitaev magnetism}},\ }\href
  {http://stacks.iop.org/0953-8984/29/i=49/a=493002} {\bibfield  {journal}
  {\bibinfo  {journal} {J. Phys. Condens. Matter}\ }\textbf {\bibinfo {volume}
  {29}},\ \bibinfo {pages} {493002} (\bibinfo {year} {2017})}\BibitemShut
  {NoStop}%
\bibitem [{\citenamefont {Wang}\ \emph {et~al.}(2019)\citenamefont {Wang},
  \citenamefont {Lu}, \citenamefont {Chen}, \citenamefont {Liu}, \citenamefont
  {Yuan}, \citenamefont {Cheong}, \citenamefont {Dong},\ and\ \citenamefont
  {Liu}}]{Sr2IrO4memory}%
  \BibitemOpen
  \bibfield  {author} {\bibinfo {author} {\bibfnamefont {H.}~\bibnamefont
  {Wang}}, \bibinfo {author} {\bibfnamefont {C.}~\bibnamefont {Lu}}, \bibinfo
  {author} {\bibfnamefont {J.}~\bibnamefont {Chen}}, \bibinfo {author}
  {\bibfnamefont {Y.}~\bibnamefont {Liu}}, \bibinfo {author} {\bibfnamefont
  {S.~L.}\ \bibnamefont {Yuan}}, \bibinfo {author} {\bibfnamefont {S.-W.}\
  \bibnamefont {Cheong}}, \bibinfo {author} {\bibfnamefont {S.}~\bibnamefont
  {Dong}},\ and\ \bibinfo {author} {\bibfnamefont {J.-M.}\ \bibnamefont
  {Liu}},\ }\bibfield  {title} {\bibinfo {title} {{Giant anisotropic
  magnetoresistance and nonvolatile memory in canted antiferromagnet
  ${\mathrm{Sr}}_{2}{\mathrm{IrO}}_{4}$}},\ }\href
  {https://doi.org/10.1038/s41467-019-10299-6} {\bibfield  {journal} {\bibinfo
  {journal} {Nature Communications}\ }\textbf {\bibinfo {volume} {10}},\
  \bibinfo {pages} {2280} (\bibinfo {year} {2019})}\BibitemShut {NoStop}%
\bibitem [{\citenamefont {Kim}\ \emph {et~al.}(2017)\citenamefont {Kim},
  \citenamefont {Jeschke}, \citenamefont {Werner},\ and\ \citenamefont
  {Valenti}}]{PhysRevLett.118.086401}%
  \BibitemOpen
  \bibfield  {author} {\bibinfo {author} {\bibfnamefont {A.~J.}\ \bibnamefont
  {Kim}}, \bibinfo {author} {\bibfnamefont {H.~O.}\ \bibnamefont {Jeschke}},
  \bibinfo {author} {\bibfnamefont {P.}~\bibnamefont {Werner}},\ and\ \bibinfo
  {author} {\bibfnamefont {R.}~\bibnamefont {Valenti}},\ }\bibfield  {title}
  {\bibinfo {title} {$\mathbf{J}$ freezing and hund's rules in
  spin-orbit-coupled multiorbital hubbard models},\ }\href
  {https://doi.org/10.1103/PhysRevLett.118.086401} {\bibfield  {journal}
  {\bibinfo  {journal} {Phys. Rev. Lett.}\ }\textbf {\bibinfo {volume} {118}},\
  \bibinfo {pages} {086401} (\bibinfo {year} {2017})}\BibitemShut {NoStop}%
\bibitem [{\citenamefont {Triebl}\ \emph {et~al.}(2018)\citenamefont {Triebl},
  \citenamefont {Kraberger}, \citenamefont {Mravlje},\ and\ \citenamefont
  {Aichhorn}}]{PhysRevB.98.205128}%
  \BibitemOpen
  \bibfield  {author} {\bibinfo {author} {\bibfnamefont {R.}~\bibnamefont
  {Triebl}}, \bibinfo {author} {\bibfnamefont {G.~J.}\ \bibnamefont
  {Kraberger}}, \bibinfo {author} {\bibfnamefont {J.}~\bibnamefont {Mravlje}},\
  and\ \bibinfo {author} {\bibfnamefont {M.}~\bibnamefont {Aichhorn}},\
  }\bibfield  {title} {\bibinfo {title} {Spin-orbit coupling and correlations
  in three-orbital systems},\ }\href
  {https://doi.org/10.1103/PhysRevB.98.205128} {\bibfield  {journal} {\bibinfo
  {journal} {Phys. Rev. B}\ }\textbf {\bibinfo {volume} {98}},\ \bibinfo
  {pages} {205128} (\bibinfo {year} {2018})}\BibitemShut {NoStop}%
\bibitem [{\citenamefont {Pajskr}\ \emph {et~al.}(2016)\citenamefont {Pajskr},
  \citenamefont {Nov\'ak}, \citenamefont {Pokorn\'y}, \citenamefont
  {Koloren\ifmmode~\check{c}\else \v{c}\fi{}}, \citenamefont {Arita},\ and\
  \citenamefont {Kune\ifmmode~\check{s}\else \v{s}\fi{}}}]{PhysRevB.93.035129}%
  \BibitemOpen
  \bibfield  {author} {\bibinfo {author} {\bibfnamefont {K.}~\bibnamefont
  {Pajskr}}, \bibinfo {author} {\bibfnamefont {P.}~\bibnamefont {Nov\'ak}},
  \bibinfo {author} {\bibfnamefont {V.}~\bibnamefont {Pokorn\'y}}, \bibinfo
  {author} {\bibfnamefont {J.}~\bibnamefont {Koloren\ifmmode~\check{c}\else
  \v{c}\fi{}}}, \bibinfo {author} {\bibfnamefont {R.}~\bibnamefont {Arita}},\
  and\ \bibinfo {author} {\bibfnamefont {J.}~\bibnamefont
  {Kune\ifmmode~\check{s}\else \v{s}\fi{}}},\ }\bibfield  {title} {\bibinfo
  {title} {On the possibility of excitonic magnetism in ir double
  perovskites},\ }\href {https://doi.org/10.1103/PhysRevB.93.035129} {\bibfield
   {journal} {\bibinfo  {journal} {Phys. Rev. B}\ }\textbf {\bibinfo {volume}
  {93}},\ \bibinfo {pages} {035129} (\bibinfo {year} {2016})}\BibitemShut
  {NoStop}%
\bibitem [{\citenamefont {Fuchs}\ \emph {et~al.}(2018)\citenamefont {Fuchs},
  \citenamefont {Dey}, \citenamefont {Aslan-Cansever}, \citenamefont {Maljuk},
  \citenamefont {Wurmehl}, \citenamefont {B\"uchner},\ and\ \citenamefont
  {Kataev}}]{PhysRevLett.120.237204}%
  \BibitemOpen
  \bibfield  {author} {\bibinfo {author} {\bibfnamefont {S.}~\bibnamefont
  {Fuchs}}, \bibinfo {author} {\bibfnamefont {T.}~\bibnamefont {Dey}}, \bibinfo
  {author} {\bibfnamefont {G.}~\bibnamefont {Aslan-Cansever}}, \bibinfo
  {author} {\bibfnamefont {A.}~\bibnamefont {Maljuk}}, \bibinfo {author}
  {\bibfnamefont {S.}~\bibnamefont {Wurmehl}}, \bibinfo {author} {\bibfnamefont
  {B.}~\bibnamefont {B\"uchner}},\ and\ \bibinfo {author} {\bibfnamefont
  {V.}~\bibnamefont {Kataev}},\ }\bibfield  {title} {\bibinfo {title}
  {{Unraveling the Nature of Magnetism of the $5{d}^{4}$ Double Perovskite
  ${\mathrm{Ba}}_{2}{\mathrm{YIrO}}_{6}$}},\ }\href
  {https://doi.org/10.1103/PhysRevLett.120.237204} {\bibfield  {journal}
  {\bibinfo  {journal} {Phys. Rev. Lett.}\ }\textbf {\bibinfo {volume} {120}},\
  \bibinfo {pages} {237204} (\bibinfo {year} {2018})}\BibitemShut {NoStop}%
\bibitem [{\citenamefont {Gretarsson}\ \emph
  {et~al.}(2019{\natexlab{a}})\citenamefont {Gretarsson}, \citenamefont
  {Suzuki}, \citenamefont {Kim}, \citenamefont {Ueda}, \citenamefont
  {Krautloher}, \citenamefont {Kim}, \citenamefont
  {Yava\ifmmode~\mbox{\c{s}}\else \c{s}\fi{}}, \citenamefont {Khaliullin},\
  and\ \citenamefont {Keimer}}]{RIXS_Ca2RuO4_Gretarson19}%
  \BibitemOpen
  \bibfield  {author} {\bibinfo {author} {\bibfnamefont {H.}~\bibnamefont
  {Gretarsson}}, \bibinfo {author} {\bibfnamefont {H.}~\bibnamefont {Suzuki}},
  \bibinfo {author} {\bibfnamefont {H.}~\bibnamefont {Kim}}, \bibinfo {author}
  {\bibfnamefont {K.}~\bibnamefont {Ueda}}, \bibinfo {author} {\bibfnamefont
  {M.}~\bibnamefont {Krautloher}}, \bibinfo {author} {\bibfnamefont {B.~J.}\
  \bibnamefont {Kim}}, \bibinfo {author} {\bibfnamefont {H.}~\bibnamefont
  {Yava\ifmmode~\mbox{\c{s}}\else \c{s}\fi{}}}, \bibinfo {author}
  {\bibfnamefont {G.}~\bibnamefont {Khaliullin}},\ and\ \bibinfo {author}
  {\bibfnamefont {B.}~\bibnamefont {Keimer}},\ }\bibfield  {title} {\bibinfo
  {title} {{{Observation of spin-orbit excitations and Hund's multiplets in
  {${\mathrm{Ca}}_{2}{\mathrm{RuO}}_{4}$}}}},\ }\href
  {https://doi.org/10.1103/PhysRevB.100.045123} {\bibfield  {journal} {\bibinfo
   {journal} {Phys. Rev. B}\ }\textbf {\bibinfo {volume} {100}},\ \bibinfo
  {pages} {045123} (\bibinfo {year} {2019}{\natexlab{a}})}\BibitemShut
  {NoStop}%
\bibitem [{\citenamefont {Khaliullin}(2013)}]{PhysRevLett.111.197201}%
  \BibitemOpen
  \bibfield  {author} {\bibinfo {author} {\bibfnamefont {G.}~\bibnamefont
  {Khaliullin}},\ }\bibfield  {title} {\bibinfo {title} {{Excitonic Magnetism
  in Van Vleck--type ${d}^{4}$ Mott Insulators}},\ }\href
  {https://doi.org/10.1103/PhysRevLett.111.197201} {\bibfield  {journal}
  {\bibinfo  {journal} {Phys. Rev. Lett.}\ }\textbf {\bibinfo {volume} {111}},\
  \bibinfo {pages} {197201} (\bibinfo {year} {2013})}\BibitemShut {NoStop}%
\bibitem [{\citenamefont {Anisimov}\ \emph {et~al.}(2019)\citenamefont
  {Anisimov}, \citenamefont {Aust}, \citenamefont {Khaliullin},\ and\
  \citenamefont {Daghofer}}]{PhysRevLett.122.177201}%
  \BibitemOpen
  \bibfield  {author} {\bibinfo {author} {\bibfnamefont {P.~S.}\ \bibnamefont
  {Anisimov}}, \bibinfo {author} {\bibfnamefont {F.}~\bibnamefont {Aust}},
  \bibinfo {author} {\bibfnamefont {G.}~\bibnamefont {Khaliullin}},\ and\
  \bibinfo {author} {\bibfnamefont {M.}~\bibnamefont {Daghofer}},\ }\bibfield
  {title} {\bibinfo {title} {Nontrivial triplon topology and triplon liquid in
  kitaev-heisenberg-type excitonic magnets},\ }\href
  {https://doi.org/10.1103/PhysRevLett.122.177201} {\bibfield  {journal}
  {\bibinfo  {journal} {Phys. Rev. Lett.}\ }\textbf {\bibinfo {volume} {122}},\
  \bibinfo {pages} {177201} (\bibinfo {year} {2019})}\BibitemShut {NoStop}%
\bibitem [{\citenamefont {Chaloupka}\ and\ \citenamefont
  {Khaliullin}(2019)}]{PhysRevB.100.224413}%
  \BibitemOpen
  \bibfield  {author} {\bibinfo {author} {\bibfnamefont {J.}~\bibnamefont
  {Chaloupka}}\ and\ \bibinfo {author} {\bibfnamefont {G.}~\bibnamefont
  {Khaliullin}},\ }\bibfield  {title} {\bibinfo {title} {Highly frustrated
  magnetism in relativistic ${d}^{4}$ mott insulators: Bosonic analog of the
  kitaev honeycomb model},\ }\href
  {https://doi.org/10.1103/PhysRevB.100.224413} {\bibfield  {journal} {\bibinfo
   {journal} {Phys. Rev. B}\ }\textbf {\bibinfo {volume} {100}},\ \bibinfo
  {pages} {224413} (\bibinfo {year} {2019})}\BibitemShut {NoStop}%
\bibitem [{\citenamefont {Jain}\ \emph {et~al.}(2017)\citenamefont {Jain},
  \citenamefont {Krautloher}, \citenamefont {Porras}, \citenamefont {Ryu},
  \citenamefont {Chen}, \citenamefont {Abernathy}, \citenamefont {Park},
  \citenamefont {Ivanov}, \citenamefont {Chaloupka}, \citenamefont
  {Khaliullin}, \citenamefont {Keimer},\ and\ \citenamefont {Kim}}]{Higgs_Ru}%
  \BibitemOpen
  \bibfield  {author} {\bibinfo {author} {\bibfnamefont {A.}~\bibnamefont
  {Jain}}, \bibinfo {author} {\bibfnamefont {M.}~\bibnamefont {Krautloher}},
  \bibinfo {author} {\bibfnamefont {J.}~\bibnamefont {Porras}}, \bibinfo
  {author} {\bibfnamefont {G.~H.}\ \bibnamefont {Ryu}}, \bibinfo {author}
  {\bibfnamefont {D.~P.}\ \bibnamefont {Chen}}, \bibinfo {author}
  {\bibfnamefont {D.~L.}\ \bibnamefont {Abernathy}}, \bibinfo {author}
  {\bibfnamefont {J.~T.}\ \bibnamefont {Park}}, \bibinfo {author}
  {\bibfnamefont {A.}~\bibnamefont {Ivanov}}, \bibinfo {author} {\bibfnamefont
  {J.}~\bibnamefont {Chaloupka}}, \bibinfo {author} {\bibfnamefont
  {G.}~\bibnamefont {Khaliullin}}, \bibinfo {author} {\bibfnamefont
  {B.}~\bibnamefont {Keimer}},\ and\ \bibinfo {author} {\bibfnamefont {B.~J.}\
  \bibnamefont {Kim}},\ }\bibfield  {title} {\bibinfo {title} {{Higgs mode and
  its decay in a two-dimensional antiferromagnet}},\ }\href
  {https://doi.org/10.1038/nphys4077} {\bibfield  {journal} {\bibinfo
  {journal} {Nature Physics}\ }\textbf {\bibinfo {volume} {13}},\ \bibinfo
  {pages} {633} (\bibinfo {year} {2017})}\BibitemShut {NoStop}%
\bibitem [{\citenamefont {Kaushal}\ \emph {et~al.}(2017)\citenamefont
  {Kaushal}, \citenamefont {Herbrych}, \citenamefont {Nocera}, \citenamefont
  {Alvarez}, \citenamefont {Moreo}, \citenamefont {Reboredo},\ and\
  \citenamefont {Dagotto}}]{PhysRevB.96.155111}%
  \BibitemOpen
  \bibfield  {author} {\bibinfo {author} {\bibfnamefont {N.}~\bibnamefont
  {Kaushal}}, \bibinfo {author} {\bibfnamefont {J.}~\bibnamefont {Herbrych}},
  \bibinfo {author} {\bibfnamefont {A.}~\bibnamefont {Nocera}}, \bibinfo
  {author} {\bibfnamefont {G.}~\bibnamefont {Alvarez}}, \bibinfo {author}
  {\bibfnamefont {A.}~\bibnamefont {Moreo}}, \bibinfo {author} {\bibfnamefont
  {F.~A.}\ \bibnamefont {Reboredo}},\ and\ \bibinfo {author} {\bibfnamefont
  {E.}~\bibnamefont {Dagotto}},\ }\bibfield  {title} {\bibinfo {title} {Density
  matrix renormalization group study of a three-orbital hubbard model with
  spin-orbit coupling in one dimension},\ }\href
  {https://doi.org/10.1103/PhysRevB.96.155111} {\bibfield  {journal} {\bibinfo
  {journal} {Phys. Rev. B}\ }\textbf {\bibinfo {volume} {96}},\ \bibinfo
  {pages} {155111} (\bibinfo {year} {2017})}\BibitemShut {NoStop}%
\bibitem [{\citenamefont {Kaushal}\ \emph {et~al.}(2020)\citenamefont
  {Kaushal}, \citenamefont {Soni}, \citenamefont {Nocera}, \citenamefont
  {Alvarez},\ and\ \citenamefont {Dagotto}}]{PhysRevB.101.245147}%
  \BibitemOpen
  \bibfield  {author} {\bibinfo {author} {\bibfnamefont {N.}~\bibnamefont
  {Kaushal}}, \bibinfo {author} {\bibfnamefont {R.}~\bibnamefont {Soni}},
  \bibinfo {author} {\bibfnamefont {A.}~\bibnamefont {Nocera}}, \bibinfo
  {author} {\bibfnamefont {G.}~\bibnamefont {Alvarez}},\ and\ \bibinfo {author}
  {\bibfnamefont {E.}~\bibnamefont {Dagotto}},\ }\bibfield  {title} {\bibinfo
  {title} {Bcs-bec crossover in a ${({t}_{2g})}^{4}$ excitonic magnet},\ }\href
  {https://doi.org/10.1103/PhysRevB.101.245147} {\bibfield  {journal} {\bibinfo
   {journal} {Phys. Rev. B}\ }\textbf {\bibinfo {volume} {101}},\ \bibinfo
  {pages} {245147} (\bibinfo {year} {2020})}\BibitemShut {NoStop}%
\bibitem [{\citenamefont {Sato}\ \emph {et~al.}(2019)\citenamefont {Sato},
  \citenamefont {Shirakawa},\ and\ \citenamefont
  {Yunoki}}]{PhysRevB.99.075117}%
  \BibitemOpen
  \bibfield  {author} {\bibinfo {author} {\bibfnamefont {T.}~\bibnamefont
  {Sato}}, \bibinfo {author} {\bibfnamefont {T.}~\bibnamefont {Shirakawa}},\
  and\ \bibinfo {author} {\bibfnamefont {S.}~\bibnamefont {Yunoki}},\
  }\bibfield  {title} {\bibinfo {title} {Spin-orbital entangled excitonic
  insulator with quadrupole order},\ }\href
  {https://doi.org/10.1103/PhysRevB.99.075117} {\bibfield  {journal} {\bibinfo
  {journal} {Phys. Rev. B}\ }\textbf {\bibinfo {volume} {99}},\ \bibinfo
  {pages} {075117} (\bibinfo {year} {2019})}\BibitemShut {NoStop}%
\bibitem [{\citenamefont {Souliou}\ \emph {et~al.}(2017)\citenamefont
  {Souliou}, \citenamefont {Chaloupka}, \citenamefont {Khaliullin},
  \citenamefont {Ryu}, \citenamefont {Jain}, \citenamefont {Kim}, \citenamefont
  {Le~Tacon},\ and\ \citenamefont {Keimer}}]{PhysRevLett.119.067201}%
  \BibitemOpen
  \bibfield  {author} {\bibinfo {author} {\bibfnamefont {S.-M.}\ \bibnamefont
  {Souliou}}, \bibinfo {author} {\bibfnamefont {J.}~\bibnamefont {Chaloupka}},
  \bibinfo {author} {\bibfnamefont {G.}~\bibnamefont {Khaliullin}}, \bibinfo
  {author} {\bibfnamefont {G.}~\bibnamefont {Ryu}}, \bibinfo {author}
  {\bibfnamefont {A.}~\bibnamefont {Jain}}, \bibinfo {author} {\bibfnamefont
  {B.~J.}\ \bibnamefont {Kim}}, \bibinfo {author} {\bibfnamefont
  {M.}~\bibnamefont {Le~Tacon}},\ and\ \bibinfo {author} {\bibfnamefont
  {B.}~\bibnamefont {Keimer}},\ }\bibfield  {title} {\bibinfo {title} {{Raman
  Scattering from Higgs Mode Oscillations in the Two-Dimensional
  Antiferromagnet ${\mathrm{Ca}}_{2}{\mathrm{RuO}}_{4}$}},\ }\href
  {https://doi.org/10.1103/PhysRevLett.119.067201} {\bibfield  {journal}
  {\bibinfo  {journal} {Phys. Rev. Lett.}\ }\textbf {\bibinfo {volume} {119}},\
  \bibinfo {pages} {067201} (\bibinfo {year} {2017})}\BibitemShut {NoStop}%
\bibitem [{\citenamefont {Zegkinoglou}\ \emph {et~al.}(2005)\citenamefont
  {Zegkinoglou}, \citenamefont {Strempfer}, \citenamefont {Nelson},
  \citenamefont {Hill}, \citenamefont {Chakhalian}, \citenamefont {Bernhard},
  \citenamefont {Lang}, \citenamefont {Srajer}, \citenamefont {Fukazawa},
  \citenamefont {Nakatsuji}, \citenamefont {Maeno},\ and\ \citenamefont
  {Keimer}}]{PhysRevLett.95.136401}%
  \BibitemOpen
  \bibfield  {author} {\bibinfo {author} {\bibfnamefont {I.}~\bibnamefont
  {Zegkinoglou}}, \bibinfo {author} {\bibfnamefont {J.}~\bibnamefont
  {Strempfer}}, \bibinfo {author} {\bibfnamefont {C.~S.}\ \bibnamefont
  {Nelson}}, \bibinfo {author} {\bibfnamefont {J.~P.}\ \bibnamefont {Hill}},
  \bibinfo {author} {\bibfnamefont {J.}~\bibnamefont {Chakhalian}}, \bibinfo
  {author} {\bibfnamefont {C.}~\bibnamefont {Bernhard}}, \bibinfo {author}
  {\bibfnamefont {J.~C.}\ \bibnamefont {Lang}}, \bibinfo {author}
  {\bibfnamefont {G.}~\bibnamefont {Srajer}}, \bibinfo {author} {\bibfnamefont
  {H.}~\bibnamefont {Fukazawa}}, \bibinfo {author} {\bibfnamefont
  {S.}~\bibnamefont {Nakatsuji}}, \bibinfo {author} {\bibfnamefont
  {Y.}~\bibnamefont {Maeno}},\ and\ \bibinfo {author} {\bibfnamefont
  {B.}~\bibnamefont {Keimer}},\ }\bibfield  {title} {\bibinfo {title} {{Orbital
  Ordering Transition in ${\mathrm{Ca}}_{2}{\mathrm{RuO}}_{4}$ Observed with
  Resonant X-Ray Diffraction}},\ }\href
  {https://doi.org/10.1103/PhysRevLett.95.136401} {\bibfield  {journal}
  {\bibinfo  {journal} {Phys. Rev. Lett.}\ }\textbf {\bibinfo {volume} {95}},\
  \bibinfo {pages} {136401} (\bibinfo {year} {2005})}\BibitemShut {NoStop}%
\bibitem [{\citenamefont {Mizokawa}\ \emph {et~al.}(2001)\citenamefont
  {Mizokawa}, \citenamefont {Tjeng}, \citenamefont {Sawatzky}, \citenamefont
  {Ghiringhelli}, \citenamefont {Tjernberg}, \citenamefont {Brookes},
  \citenamefont {Fukazawa}, \citenamefont {Nakatsuji},\ and\ \citenamefont
  {Maeno}}]{PhysRevLett.87.077202}%
  \BibitemOpen
  \bibfield  {author} {\bibinfo {author} {\bibfnamefont {T.}~\bibnamefont
  {Mizokawa}}, \bibinfo {author} {\bibfnamefont {L.~H.}\ \bibnamefont {Tjeng}},
  \bibinfo {author} {\bibfnamefont {G.~A.}\ \bibnamefont {Sawatzky}}, \bibinfo
  {author} {\bibfnamefont {G.}~\bibnamefont {Ghiringhelli}}, \bibinfo {author}
  {\bibfnamefont {O.}~\bibnamefont {Tjernberg}}, \bibinfo {author}
  {\bibfnamefont {N.~B.}\ \bibnamefont {Brookes}}, \bibinfo {author}
  {\bibfnamefont {H.}~\bibnamefont {Fukazawa}}, \bibinfo {author}
  {\bibfnamefont {S.}~\bibnamefont {Nakatsuji}},\ and\ \bibinfo {author}
  {\bibfnamefont {Y.}~\bibnamefont {Maeno}},\ }\bibfield  {title} {\bibinfo
  {title} {{Spin-Orbit Coupling in the Mott Insulator
  ${\mathrm{Ca}}_{2}{\mathrm{RuO}}_{4}$}},\ }\href
  {https://doi.org/10.1103/PhysRevLett.87.077202} {\bibfield  {journal}
  {\bibinfo  {journal} {Phys. Rev. Lett.}\ }\textbf {\bibinfo {volume} {87}},\
  \bibinfo {pages} {077202} (\bibinfo {year} {2001})}\BibitemShut {NoStop}%
\bibitem [{\citenamefont {Kunkem\"oller}\ \emph {et~al.}(2015)\citenamefont
  {Kunkem\"oller}, \citenamefont {Khomskii}, \citenamefont {Steffens},
  \citenamefont {Piovano}, \citenamefont {Nugroho},\ and\ \citenamefont
  {Braden}}]{PhysRevLett.115.247201}%
  \BibitemOpen
  \bibfield  {author} {\bibinfo {author} {\bibfnamefont {S.}~\bibnamefont
  {Kunkem\"oller}}, \bibinfo {author} {\bibfnamefont {D.}~\bibnamefont
  {Khomskii}}, \bibinfo {author} {\bibfnamefont {P.}~\bibnamefont {Steffens}},
  \bibinfo {author} {\bibfnamefont {A.}~\bibnamefont {Piovano}}, \bibinfo
  {author} {\bibfnamefont {A.~A.}\ \bibnamefont {Nugroho}},\ and\ \bibinfo
  {author} {\bibfnamefont {M.}~\bibnamefont {Braden}},\ }\bibfield  {title}
  {\bibinfo {title} {{Highly Anisotropic Magnon Dispersion in
  ${\mathrm{Ca}}_{2}{\mathrm{RuO}}_{4}$: Evidence for Strong Spin Orbit
  Coupling}},\ }\href {https://doi.org/10.1103/PhysRevLett.115.247201}
  {\bibfield  {journal} {\bibinfo  {journal} {Phys. Rev. Lett.}\ }\textbf
  {\bibinfo {volume} {115}},\ \bibinfo {pages} {247201} (\bibinfo {year}
  {2015})}\BibitemShut {NoStop}%
\bibitem [{\citenamefont {Zhang}\ and\ \citenamefont
  {Pavarini}(2020)}]{PhysRevB.101.205128}%
  \BibitemOpen
  \bibfield  {author} {\bibinfo {author} {\bibfnamefont {G.}~\bibnamefont
  {Zhang}}\ and\ \bibinfo {author} {\bibfnamefont {E.}~\bibnamefont
  {Pavarini}},\ }\bibfield  {title} {\bibinfo {title} {{Higgs mode and
  stability of $xy$-orbital ordering in
  ${\mathrm{Ca}}_{2}{\mathrm{RuO}}_{4}$}},\ }\href
  {https://doi.org/10.1103/PhysRevB.101.205128} {\bibfield  {journal} {\bibinfo
   {journal} {Phys. Rev. B}\ }\textbf {\bibinfo {volume} {101}},\ \bibinfo
  {pages} {205128} (\bibinfo {year} {2020})}\BibitemShut {NoStop}%
\bibitem [{\citenamefont {Feldmaier}\ \emph {et~al.}(2020)\citenamefont
  {Feldmaier}, \citenamefont {Strobel}, \citenamefont {Schmid}, \citenamefont
  {Hansmann},\ and\ \citenamefont {Daghofer}}]{PhysRevResearch.2.033201}%
  \BibitemOpen
  \bibfield  {author} {\bibinfo {author} {\bibfnamefont {T.}~\bibnamefont
  {Feldmaier}}, \bibinfo {author} {\bibfnamefont {P.}~\bibnamefont {Strobel}},
  \bibinfo {author} {\bibfnamefont {M.}~\bibnamefont {Schmid}}, \bibinfo
  {author} {\bibfnamefont {P.}~\bibnamefont {Hansmann}},\ and\ \bibinfo
  {author} {\bibfnamefont {M.}~\bibnamefont {Daghofer}},\ }\bibfield  {title}
  {\bibinfo {title} {{Excitonic magnetism at the intersection of spin-orbit
  coupling and crystal-field splitting}},\ }\href
  {https://doi.org/10.1103/PhysRevResearch.2.033201} {\bibfield  {journal}
  {\bibinfo  {journal} {Phys. Rev. Research}\ }\textbf {\bibinfo {volume}
  {2}},\ \bibinfo {pages} {033201} (\bibinfo {year} {2020})}\BibitemShut
  {NoStop}%
\bibitem [{\citenamefont {{Lotze}}\ and\ \citenamefont
  {{Daghofer}}(2021)}]{2021arXiv210205489L}%
  \BibitemOpen
  \bibfield  {author} {\bibinfo {author} {\bibfnamefont {J.}~\bibnamefont
  {{Lotze}}}\ and\ \bibinfo {author} {\bibfnamefont {M.}~\bibnamefont
  {{Daghofer}}},\ }\bibfield  {title} {\bibinfo {title} {{Suppression of
  effective spin-orbit coupling by thermal fluctuations in spin-orbit coupled
  antiferromagnets}},\ }\href@noop {} {\bibfield  {journal} {\bibinfo
  {journal} {arXiv e-prints}\ ,\ \bibinfo {eid} {arXiv:2102.05489}} (\bibinfo
  {year} {2021})},\ \Eprint {https://arxiv.org/abs/2102.05489}
  {arXiv:2102.05489 [cond-mat.str-el]} \BibitemShut {NoStop}%
\bibitem [{\citenamefont {Mohapatra}\ and\ \citenamefont
  {Singh}(2020)}]{Mohapatra_2020}%
  \BibitemOpen
  \bibfield  {author} {\bibinfo {author} {\bibfnamefont {S.}~\bibnamefont
  {Mohapatra}}\ and\ \bibinfo {author} {\bibfnamefont {A.}~\bibnamefont
  {Singh}},\ }\bibfield  {title} {\bibinfo {title} {Magnetic reorientation
  transition in a three orbital model for ca2ruo4{\textemdash}interplay of
  spin{\textendash}orbit coupling, tetragonal distortion, and coulomb
  interactions},\ }\href {https://doi.org/10.1088/1361-648x/abacad} {\bibfield
  {journal} {\bibinfo  {journal} {Journal of Physics: Condensed Matter}\
  }\textbf {\bibinfo {volume} {32}},\ \bibinfo {pages} {485805} (\bibinfo
  {year} {2020})}\BibitemShut {NoStop}%
\bibitem [{\citenamefont {Svoboda}\ \emph {et~al.}(2017)\citenamefont
  {Svoboda}, \citenamefont {Randeria},\ and\ \citenamefont
  {Trivedi}}]{PhysRevB.95.014409}%
  \BibitemOpen
  \bibfield  {author} {\bibinfo {author} {\bibfnamefont {C.}~\bibnamefont
  {Svoboda}}, \bibinfo {author} {\bibfnamefont {M.}~\bibnamefont {Randeria}},\
  and\ \bibinfo {author} {\bibfnamefont {N.}~\bibnamefont {Trivedi}},\
  }\bibfield  {title} {\bibinfo {title} {Effective magnetic interactions in
  spin-orbit coupled ${d}^{4}$ mott insulators},\ }\href
  {https://doi.org/10.1103/PhysRevB.95.014409} {\bibfield  {journal} {\bibinfo
  {journal} {Phys. Rev. B}\ }\textbf {\bibinfo {volume} {95}},\ \bibinfo
  {pages} {014409} (\bibinfo {year} {2017})}\BibitemShut {NoStop}%
\bibitem [{\citenamefont {Ole\'{s}}(1983)}]{PhysRevB.28.327}%
  \BibitemOpen
  \bibfield  {author} {\bibinfo {author} {\bibfnamefont {A.~M.}\ \bibnamefont
  {Ole\'{s}}},\ }\bibfield  {title} {\bibinfo {title} {{Antiferromagnetism and
  correlation of electrons in transition metals}},\ }\href
  {https://doi.org/10.1103/PhysRevB.28.327} {\bibfield  {journal} {\bibinfo
  {journal} {Phys. Rev. B}\ }\textbf {\bibinfo {volume} {28}},\ \bibinfo
  {pages} {327} (\bibinfo {year} {1983})}\BibitemShut {NoStop}%
\bibitem [{\citenamefont {Chaloupka}\ \emph
  {et~al.}(2010{\natexlab{b}})\citenamefont {Chaloupka}, \citenamefont
  {Jackeli},\ and\ \citenamefont {Khaliullin}}]{PhysRevLett.105.027204}%
  \BibitemOpen
  \bibfield  {author} {\bibinfo {author} {\bibfnamefont {J.}~\bibnamefont
  {Chaloupka}}, \bibinfo {author} {\bibfnamefont {G.}~\bibnamefont {Jackeli}},\
  and\ \bibinfo {author} {\bibfnamefont {G.}~\bibnamefont {Khaliullin}},\
  }\bibfield  {title} {\bibinfo {title} {{Kitaev-Heisenberg Model on a
  Honeycomb Lattice: Possible Exotic Phases in Iridium Oxides
  ${A}_{2}{\mathrm{IrO}}_{3}$}},\ }\href
  {https://doi.org/10.1103/PhysRevLett.105.027204} {\bibfield  {journal}
  {\bibinfo  {journal} {Phys. Rev. Lett.}\ }\textbf {\bibinfo {volume} {105}},\
  \bibinfo {pages} {027204} (\bibinfo {year} {2010}{\natexlab{b}})}\BibitemShut
  {NoStop}%
\bibitem [{\citenamefont {Streltsov}\ and\ \citenamefont
  {Khomskii}(2017)}]{Streltsov}%
  \BibitemOpen
  \bibfield  {author} {\bibinfo {author} {\bibfnamefont {S.~V.}\ \bibnamefont
  {Streltsov}}\ and\ \bibinfo {author} {\bibfnamefont {D.~I.}\ \bibnamefont
  {Khomskii}},\ }\bibfield  {title} {\bibinfo {title} {{Orbital physics in
  transition metal compounds: new trends}},\ }\href
  {https://doi.org/10.3367/ufne.2017.08.038196} {\bibfield  {journal} {\bibinfo
   {journal} {Physics-Uspekhi}\ }\textbf {\bibinfo {volume} {60}},\ \bibinfo
  {pages} {1121} (\bibinfo {year} {2017})}\BibitemShut {NoStop}%
\bibitem [{\citenamefont {Kugel}\ and\ \citenamefont {Khomskii}(1982)}]{Kugel}%
  \BibitemOpen
  \bibfield  {author} {\bibinfo {author} {\bibfnamefont {K.~I.}\ \bibnamefont
  {Kugel}}\ and\ \bibinfo {author} {\bibfnamefont {D.~I.}\ \bibnamefont
  {Khomskii}},\ }\bibfield  {title} {\bibinfo {title} {{The Jahn-Teller effect
  and magnetism: transition metal compounds}},\ }\href
  {https://doi.org/10.1070/pu1982v025n04abeh004537} {\bibfield  {journal}
  {\bibinfo  {journal} {Soviet Physics Uspekhi}\ }\textbf {\bibinfo {volume}
  {25}},\ \bibinfo {pages} {231} (\bibinfo {year} {1982})}\BibitemShut
  {NoStop}%
\bibitem [{\citenamefont {Cuoco}\ \emph
  {et~al.}(2006{\natexlab{a}})\citenamefont {Cuoco}, \citenamefont {Forte},\
  and\ \citenamefont {Noce}}]{PhysRevB.73.094428}%
  \BibitemOpen
  \bibfield  {author} {\bibinfo {author} {\bibfnamefont {M.}~\bibnamefont
  {Cuoco}}, \bibinfo {author} {\bibfnamefont {F.}~\bibnamefont {Forte}},\ and\
  \bibinfo {author} {\bibfnamefont {C.}~\bibnamefont {Noce}},\ }\bibfield
  {title} {\bibinfo {title} {Probing spin-orbital-lattice correlations in
  $4{d}^{4}$ systems},\ }\href {https://doi.org/10.1103/PhysRevB.73.094428}
  {\bibfield  {journal} {\bibinfo  {journal} {Phys. Rev. B}\ }\textbf {\bibinfo
  {volume} {73}},\ \bibinfo {pages} {094428} (\bibinfo {year}
  {2006}{\natexlab{a}})}\BibitemShut {NoStop}%
\bibitem [{\citenamefont {Akbari}\ and\ \citenamefont
  {Khaliullin}(2014)}]{PhysRevB.90.035137}%
  \BibitemOpen
  \bibfield  {author} {\bibinfo {author} {\bibfnamefont {A.}~\bibnamefont
  {Akbari}}\ and\ \bibinfo {author} {\bibfnamefont {G.}~\bibnamefont
  {Khaliullin}},\ }\bibfield  {title} {\bibinfo {title} {{Magnetic excitations
  in a spin-orbit-coupled ${d}^{4}$ Mott insulator on the square lattice}},\
  }\href {https://doi.org/10.1103/PhysRevB.90.035137} {\bibfield  {journal}
  {\bibinfo  {journal} {Phys. Rev. B}\ }\textbf {\bibinfo {volume} {90}},\
  \bibinfo {pages} {035137} (\bibinfo {year} {2014})}\BibitemShut {NoStop}%
\bibitem [{\citenamefont {Jackeli}\ and\ \citenamefont
  {Khaliullin}(2009)}]{PhysRevLett.102.017205}%
  \BibitemOpen
  \bibfield  {author} {\bibinfo {author} {\bibfnamefont {G.}~\bibnamefont
  {Jackeli}}\ and\ \bibinfo {author} {\bibfnamefont {G.}~\bibnamefont
  {Khaliullin}},\ }\bibfield  {title} {\bibinfo {title} {{Mott Insulators in
  the Strong Spin-Orbit Coupling Limit: From Heisenberg to a Quantum Compass
  and Kitaev Models}},\ }\href {https://doi.org/10.1103/PhysRevLett.102.017205}
  {\bibfield  {journal} {\bibinfo  {journal} {Phys. Rev. Lett.}\ }\textbf
  {\bibinfo {volume} {102}},\ \bibinfo {pages} {017205} (\bibinfo {year}
  {2009})}\BibitemShut {NoStop}%
\bibitem [{\citenamefont {Bertinshaw}\ \emph
  {et~al.}(2019{\natexlab{b}})\citenamefont {Bertinshaw}, \citenamefont
  {Gurung}, \citenamefont {Jorba}, \citenamefont {Liu}, \citenamefont {Schmid},
  \citenamefont {Mantadakis}, \citenamefont {Daghofer}, \citenamefont
  {Krautloher}, \citenamefont {Jain}, \citenamefont {Ryu}, \citenamefont
  {Fabelo}, \citenamefont {Hansmann}, \citenamefont {Khaliullin}, \citenamefont
  {Pfleiderer}, \citenamefont {Keimer},\ and\ \citenamefont
  {Kim}}]{PhysRevLett.123.137204}%
  \BibitemOpen
  \bibfield  {author} {\bibinfo {author} {\bibfnamefont {J.}~\bibnamefont
  {Bertinshaw}}, \bibinfo {author} {\bibfnamefont {N.}~\bibnamefont {Gurung}},
  \bibinfo {author} {\bibfnamefont {P.}~\bibnamefont {Jorba}}, \bibinfo
  {author} {\bibfnamefont {H.}~\bibnamefont {Liu}}, \bibinfo {author}
  {\bibfnamefont {M.}~\bibnamefont {Schmid}}, \bibinfo {author} {\bibfnamefont
  {D.~T.}\ \bibnamefont {Mantadakis}}, \bibinfo {author} {\bibfnamefont
  {M.}~\bibnamefont {Daghofer}}, \bibinfo {author} {\bibfnamefont
  {M.}~\bibnamefont {Krautloher}}, \bibinfo {author} {\bibfnamefont
  {A.}~\bibnamefont {Jain}}, \bibinfo {author} {\bibfnamefont {G.~H.}\
  \bibnamefont {Ryu}}, \bibinfo {author} {\bibfnamefont {O.}~\bibnamefont
  {Fabelo}}, \bibinfo {author} {\bibfnamefont {P.}~\bibnamefont {Hansmann}},
  \bibinfo {author} {\bibfnamefont {G.}~\bibnamefont {Khaliullin}}, \bibinfo
  {author} {\bibfnamefont {C.}~\bibnamefont {Pfleiderer}}, \bibinfo {author}
  {\bibfnamefont {B.}~\bibnamefont {Keimer}},\ and\ \bibinfo {author}
  {\bibfnamefont {B.~J.}\ \bibnamefont {Kim}},\ }\bibfield  {title} {\bibinfo
  {title} {{Unique Crystal Structure of ${\mathrm{Ca}}_{2}{\mathrm{RuO}}_{4}$
  in the Current Stabilized Semimetallic State}},\ }\href
  {https://doi.org/10.1103/PhysRevLett.123.137204} {\bibfield  {journal}
  {\bibinfo  {journal} {Phys. Rev. Lett.}\ }\textbf {\bibinfo {volume} {123}},\
  \bibinfo {pages} {137204} (\bibinfo {year} {2019}{\natexlab{b}})}\BibitemShut
  {NoStop}%
\bibitem [{\citenamefont {Gretarsson}\ \emph
  {et~al.}(2019{\natexlab{b}})\citenamefont {Gretarsson}, \citenamefont
  {Suzuki}, \citenamefont {Kim}, \citenamefont {Ueda}, \citenamefont
  {Krautloher}, \citenamefont {Kim}, \citenamefont
  {Yava\ifmmode~\mbox{\c{s}}\else \c{s}\fi{}}, \citenamefont {Khaliullin},\
  and\ \citenamefont {Keimer}}]{PhysRevB.100.045123}%
  \BibitemOpen
  \bibfield  {author} {\bibinfo {author} {\bibfnamefont {H.}~\bibnamefont
  {Gretarsson}}, \bibinfo {author} {\bibfnamefont {H.}~\bibnamefont {Suzuki}},
  \bibinfo {author} {\bibfnamefont {H.}~\bibnamefont {Kim}}, \bibinfo {author}
  {\bibfnamefont {K.}~\bibnamefont {Ueda}}, \bibinfo {author} {\bibfnamefont
  {M.}~\bibnamefont {Krautloher}}, \bibinfo {author} {\bibfnamefont {B.~J.}\
  \bibnamefont {Kim}}, \bibinfo {author} {\bibfnamefont {H.}~\bibnamefont
  {Yava\ifmmode~\mbox{\c{s}}\else \c{s}\fi{}}}, \bibinfo {author}
  {\bibfnamefont {G.}~\bibnamefont {Khaliullin}},\ and\ \bibinfo {author}
  {\bibfnamefont {B.}~\bibnamefont {Keimer}},\ }\bibfield  {title} {\bibinfo
  {title} {{Observation of spin-orbit excitations and Hund's multiplets in
  ${\mathrm{Ca}}_{2}{\mathrm{RuO}}_{4}$}},\ }\href
  {https://doi.org/10.1103/PhysRevB.100.045123} {\bibfield  {journal} {\bibinfo
   {journal} {Phys. Rev. B}\ }\textbf {\bibinfo {volume} {100}},\ \bibinfo
  {pages} {045123} (\bibinfo {year} {2019}{\natexlab{b}})}\BibitemShut
  {NoStop}%
\bibitem [{\citenamefont {Stoudenmire}\ \emph {et~al.}(2009)\citenamefont
  {Stoudenmire}, \citenamefont {Trebst},\ and\ \citenamefont
  {Balents}}]{PhysRevB.79.214436}%
  \BibitemOpen
  \bibfield  {author} {\bibinfo {author} {\bibfnamefont {E.~M.}\ \bibnamefont
  {Stoudenmire}}, \bibinfo {author} {\bibfnamefont {S.}~\bibnamefont
  {Trebst}},\ and\ \bibinfo {author} {\bibfnamefont {L.}~\bibnamefont
  {Balents}},\ }\bibfield  {title} {\bibinfo {title} {Quadrupolar correlations
  and spin freezing in $s=1$ triangular lattice antiferromagnets},\ }\href
  {https://doi.org/10.1103/PhysRevB.79.214436} {\bibfield  {journal} {\bibinfo
  {journal} {Phys. Rev. B}\ }\textbf {\bibinfo {volume} {79}},\ \bibinfo
  {pages} {214436} (\bibinfo {year} {2009})}\BibitemShut {NoStop}%
\bibitem [{\citenamefont {Cuoco}\ \emph
  {et~al.}(2006{\natexlab{b}})\citenamefont {Cuoco}, \citenamefont {Forte},\
  and\ \citenamefont {Noce}}]{PhysRevB.74.195124}%
  \BibitemOpen
  \bibfield  {author} {\bibinfo {author} {\bibfnamefont {M.}~\bibnamefont
  {Cuoco}}, \bibinfo {author} {\bibfnamefont {F.}~\bibnamefont {Forte}},\ and\
  \bibinfo {author} {\bibfnamefont {C.}~\bibnamefont {Noce}},\ }\bibfield
  {title} {\bibinfo {title} {{Interplay of Coulomb interactions and $c$-axis
  octahedra distortions in single-layer ruthenates}},\ }\href
  {https://doi.org/10.1103/PhysRevB.74.195124} {\bibfield  {journal} {\bibinfo
  {journal} {Phys. Rev. B}\ }\textbf {\bibinfo {volume} {74}},\ \bibinfo
  {pages} {195124} (\bibinfo {year} {2006}{\natexlab{b}})}\BibitemShut
  {NoStop}%
\bibitem [{\citenamefont {Khaliullin}\ \emph {et~al.}(2001)\citenamefont
  {Khaliullin}, \citenamefont {Horsch},\ and\ \citenamefont
  {Ole\ifmmode~\acute{s}\else \'{s}\fi{}}}]{PhysRevLett.86.3879}%
  \BibitemOpen
  \bibfield  {author} {\bibinfo {author} {\bibfnamefont {G.}~\bibnamefont
  {Khaliullin}}, \bibinfo {author} {\bibfnamefont {P.}~\bibnamefont {Horsch}},\
  and\ \bibinfo {author} {\bibfnamefont {A.~M.}\ \bibnamefont
  {Ole\ifmmode~\acute{s}\else \'{s}\fi{}}},\ }\bibfield  {title} {\bibinfo
  {title} {Spin order due to orbital fluctuations: Cubic vanadates},\ }\href
  {https://doi.org/10.1103/PhysRevLett.86.3879} {\bibfield  {journal} {\bibinfo
   {journal} {Phys. Rev. Lett.}\ }\textbf {\bibinfo {volume} {86}},\ \bibinfo
  {pages} {3879} (\bibinfo {year} {2001})}\BibitemShut {NoStop}%
\bibitem [{\citenamefont {Hotta}\ and\ \citenamefont
  {Dagotto}(2001)}]{PhysRevLett.88.017201}%
  \BibitemOpen
  \bibfield  {author} {\bibinfo {author} {\bibfnamefont {T.}~\bibnamefont
  {Hotta}}\ and\ \bibinfo {author} {\bibfnamefont {E.}~\bibnamefont
  {Dagotto}},\ }\bibfield  {title} {\bibinfo {title} {Prediction of orbital
  ordering in single-layered ruthenates},\ }\href
  {https://doi.org/10.1103/PhysRevLett.88.017201} {\bibfield  {journal}
  {\bibinfo  {journal} {Phys. Rev. Lett.}\ }\textbf {\bibinfo {volume} {88}},\
  \bibinfo {pages} {017201} (\bibinfo {year} {2001})}\BibitemShut {NoStop}%
\bibitem [{\citenamefont {Zhang}\ and\ \citenamefont
  {Pavarini}(2017)}]{PhysRevB.95.075145}%
  \BibitemOpen
  \bibfield  {author} {\bibinfo {author} {\bibfnamefont {G.}~\bibnamefont
  {Zhang}}\ and\ \bibinfo {author} {\bibfnamefont {E.}~\bibnamefont
  {Pavarini}},\ }\bibfield  {title} {\bibinfo {title} {{Mott transition,
  spin-orbit effects, and magnetism in
  ${\mathrm{Ca}}_{2}{\mathrm{RuO}}_{4}$}},\ }\href
  {https://doi.org/10.1103/PhysRevB.95.075145} {\bibfield  {journal} {\bibinfo
  {journal} {Phys. Rev. B}\ }\textbf {\bibinfo {volume} {95}},\ \bibinfo
  {pages} {075145} (\bibinfo {year} {2017})}\BibitemShut {NoStop}%
\bibitem [{\citenamefont {Sutter}\ \emph {et~al.}(2017)\citenamefont {Sutter},
  \citenamefont {Fatuzzo}, \citenamefont {Moser}, \citenamefont {Kim},
  \citenamefont {Fittipaldi}, \citenamefont {Vecchione}, \citenamefont
  {Granata}, \citenamefont {Sassa}, \citenamefont {Cossalter}, \citenamefont
  {Gatti}, \citenamefont {Grioni}, \citenamefont {Rønnow}, \citenamefont
  {Plumb}, \citenamefont {Matt}, \citenamefont {Shi}, \citenamefont {Hoesch},
  \citenamefont {Kim}, \citenamefont {Chang}, \citenamefont {Jeng},
  \citenamefont {Jozwiak}, \citenamefont {Bostwick}, \citenamefont {Rotenberg},
  \citenamefont {Georges}, \citenamefont {Neupert},\ and\ \citenamefont
  {Chang}}]{hund_Ca2RuO4}%
  \BibitemOpen
  \bibfield  {author} {\bibinfo {author} {\bibfnamefont {D.}~\bibnamefont
  {Sutter}}, \bibinfo {author} {\bibfnamefont {C.~G.}\ \bibnamefont {Fatuzzo}},
  \bibinfo {author} {\bibfnamefont {S.}~\bibnamefont {Moser}}, \bibinfo
  {author} {\bibfnamefont {M.}~\bibnamefont {Kim}}, \bibinfo {author}
  {\bibfnamefont {R.}~\bibnamefont {Fittipaldi}}, \bibinfo {author}
  {\bibfnamefont {A.}~\bibnamefont {Vecchione}}, \bibinfo {author}
  {\bibfnamefont {V.}~\bibnamefont {Granata}}, \bibinfo {author} {\bibfnamefont
  {Y.}~\bibnamefont {Sassa}}, \bibinfo {author} {\bibfnamefont
  {F.}~\bibnamefont {Cossalter}}, \bibinfo {author} {\bibfnamefont
  {G.}~\bibnamefont {Gatti}}, \bibinfo {author} {\bibfnamefont
  {M.}~\bibnamefont {Grioni}}, \bibinfo {author} {\bibfnamefont {H.~M.}\
  \bibnamefont {Rønnow}}, \bibinfo {author} {\bibfnamefont {N.~C.}\
  \bibnamefont {Plumb}}, \bibinfo {author} {\bibfnamefont {C.~E.}\ \bibnamefont
  {Matt}}, \bibinfo {author} {\bibfnamefont {M.}~\bibnamefont {Shi}}, \bibinfo
  {author} {\bibfnamefont {M.}~\bibnamefont {Hoesch}}, \bibinfo {author}
  {\bibfnamefont {T.~K.}\ \bibnamefont {Kim}}, \bibinfo {author} {\bibfnamefont
  {T.-R.}\ \bibnamefont {Chang}}, \bibinfo {author} {\bibfnamefont {H.-T.}\
  \bibnamefont {Jeng}}, \bibinfo {author} {\bibfnamefont {C.}~\bibnamefont
  {Jozwiak}}, \bibinfo {author} {\bibfnamefont {A.}~\bibnamefont {Bostwick}},
  \bibinfo {author} {\bibfnamefont {E.}~\bibnamefont {Rotenberg}}, \bibinfo
  {author} {\bibfnamefont {A.}~\bibnamefont {Georges}}, \bibinfo {author}
  {\bibfnamefont {T.}~\bibnamefont {Neupert}},\ and\ \bibinfo {author}
  {\bibfnamefont {J.}~\bibnamefont {Chang}},\ }\bibfield  {title} {\bibinfo
  {title} {{Hallmarks of Hunds coupling in the Mott insulator
  ${\mathrm{Ca}}_{2}{\mathrm{RuO}}_{4}$}},\ }\href
  {https://doi.org/10.1038/ncomms15176} {\bibfield  {journal} {\bibinfo
  {journal} {Nature Communications}\ }\textbf {\bibinfo {volume} {8}},\
  \bibinfo {pages} {15176} (\bibinfo {year} {2017})}\BibitemShut {NoStop}%
\end{thebibliography}%

\clearpage
\appendix

\end{document}